\documentclass[12pt]{article}
\usepackage{amsmath,amssymb,bm,epsfig}
\usepackage{dcolumn}
\usepackage{longtable}
\usepackage[format=hang]{caption}%captionの折り返し位置の調整
\newcolumntype{d}[1]{D{.}{.}{#1}}

\renewcommand{\thefootnote}{\fnsymbol{footnote}}

\begin{document}

\title{
\begin{flushright}
\begin{minipage}{0.2\linewidth}
\normalsize
WU-HEP-18-10 \\*[50pt]
\end{minipage}
\end{flushright}
{\Large \bf 
$SO(32)$ heterotic standard model vacua \\
in general Calabi-Yau compactifications
\\*[20pt]}}

\author{Hajime~Otsuka\footnote{
E-mail address: h.otsuka@aoni.waseda.jp
}
\ and\
Kenta~Takemoto\footnote{
E-mail address: phys-tm.3000@asagi.waseda.jp
}\\*[20pt]
{\it \normalsize 
Department of Physics, Waseda University, 
Tokyo 169-8555, Japan} \\*[50pt]}

\date{
\centerline{\small \bf Abstract}
\begin{minipage}{0.9\linewidth}
\medskip 
\medskip 
\small
%We explore SO(32) heterotic line bundle models with a hypercharge flux on general Calabi-Yau threefolds. 
We study a direct flux breaking scenario in $SO(32)$ heterotic string theory on general Calabi-Yau threefolds. 
%This model-building approach is attractive not only in F-theory with hypercharge flux breaking, 
%but also in heterotic string theory. 
The direct flux breaking, corresponding to hypercharge flux breaking in the F-theory context, 
allows us to derive the Standard Model in general Calabi-Yau compactifications. 
%The systematic search reveals that there exists a consistent heterotic 
%background leading to the three-generations of quarks and leptons and no chiral exotic matters. 
We present a general formula leading to the three generations of quarks and leptons and no chiral exotics 
in a background-independent way. 
As a concrete example, we show the three-generation model on a 
complete intersection Calabi-Yau threefold. 
\end{minipage}
}

\begin{titlepage}
\maketitle
\thispagestyle{empty}
%\clearpage
%\thispagestyle{empty}
\end{titlepage}

\renewcommand{\thefootnote}{\arabic{footnote}}
\setcounter{footnote}{0}
%\vspace{35pt}

\tableofcontents

%%%%%%%%%%%%%%%%%%%%%%%%%%%%%%%%%%%%%%%%%%%%%%%%%%%%%%%%%%%%%%%%%%%%%%%%%%%%%%%%%%%%%%%%%%%%%%%%%%%%%%%%%%%%%
\section{Introduction} 
String theory is an attractive candidate not only for a theory of quantum gravity but for a unified theory of elementary forces. 
It predicts phenomenologically promising higher-dimensional non-abelian gauge theories which are 
expected to include the Standard Model (SM) as well as its realistic spectra. 
Indeed, string theory naturally incorporates non-abelian gauge groups, appearing from stacks of D-branes in type I and II string theories, 
seven-branes in F-theory and closed string sector in heterotic string theories. 

Non-trivial gauge backgrounds such as internal gauge fluxes and Wilson lines play an important role in 
breaking the higher-rank gauge group down to the SM one~\cite{Witten:1984dg,Candelas:1985en}, 
but Wilson lines are only allowed for restricted non-simply-connected manifolds~\cite{Candelas:1985en}. 
As an example, there exist $195$ non-simply-connected complete intersection Calabi-Yau (CY) threefolds (CICYs) among $7820$ 
CICYs~\cite{Candelas:1987kf,Candelas:1987du,Braun:2010vc}.  
Hence, it is interesting to check whether or not gauge fluxes directly lead to the SM gauge group. 
This approach allows us to open up a new window for the string model building in more general CY compactifications. 
In the context of F-theory grand unified theories (GUTs), such a direct flux breaking called hypercharge flux breaking is an attractive scenario to break 
the $SU(5)$ gauge group~\cite{Beasley:2008kw,Donagi:2008kj} (also discussed in the dual heterotic string side~\cite{Blumenhagen:2005ga,Blumenhagen:2005pm,Blumenhagen:2006ux,Blumenhagen:2006wj}). 
In contrast, the authors of Ref.~\cite{Anderson:2014hia} pointed out that the realization of hypercharge flux scenario is generically difficult to achieve in 
$E_8\times E_8$ heterotic string theory on smooth CY threefolds due to the large number of index conditions, corresponding to the three generations of quarks and leptons. 
It therefore motivates us to search for models with a hypercharge flux in other string theories.

In this paper, we systematically study $SO(32)$ heterotic line bundle models as a realization of direct flux breaking.\footnote{See, Refs.~\cite{Abe:2015mua,Nibbelink:2015vha,Otsuka:2018oyf} for $SO(32)$ heterotic line bundle models using Wilson lines.} 
In a way similar to Ref.~\cite{Anderson:2014hia}, we search for the three-generation SM against several branchings of $SO(32)$ 
in a background-independent way. 
After solving a large number of index conditions for each elementary particle, together with the K-theory condition and hypercharge masslessness 
conditions, it turns out that the SM-like spectrum can be realized in general CY compactifications. 
Note that supersymmetric and stability conditions are required to be checked for each CY threefold. 
%it turns out that the spectrum like SU(5) and SO(10) GUTs is excluded, but it is possible 
%to obtain the Pati-Salam-like spectrum and direct standard model one. 

The remainder of this paper is as follows. 
In Sec.~\ref{sec:2}, we first show the model-building approach to derive heterotic Standard Models on smooth 
CY threefolds. After discussing the hypercharge direction for several group decompositions of $SO(32)$ and corresponding spectrum, 
we next present the general formula satisfying the index conditions, K-theory condition and hypercharge masslessness conditions. 
The obtained formula is applicable to general CY threefolds.  
In Sec.~\ref{sec:3}, the specific MSSM (minimal supersymmetric Standard Model)-like spectrum 
is shown for a concrete CICY. In Appendix A, we present the algorithm to compute the particle spectrum for all the group decompositions discussed 
in this paper. 
% and in particular we show the phenomenologically attractive models, leading to the three-generation of quarks and leptons without chiral exotics. 

%To derive the standard model gauge group from the such a non-abelian gauge theory, internal gauge flux is one of the promising tool 
%to realize not only the standard model gauge group but also the chiral spectrum in the four-dimensional effective action. 

%To decompose the higher rank gauge group such as the grand unified group  to the standard model one
%In particular, the hypercharge flux is 
%Another phenomenologically interesting tool to decompose the higher rank gauge group to the standard model one is Wilson 
%lines. can be introduced on restricted non simply-connected manifolds, it is required to obtain 
%the standard models using the only gauge flux. 

%%%%%%%%%%%%%%%%%%%%%%%%%%%%%%%%%%%%%%%%%%%%%%%%%%%%%%%%%%%%%%%%%%%%%%%%%%%%%%%%%%%%%%%%%%%%%%%%%%%%%%%%
\section{Direct flux breaking in heterotic string}
\label{sec:2}

\subsection{Model-building approach}
\label{subsec:2.1}
In the context of heterotic string, internal gauge fluxes open up wide possibilities of constructing the SM. 
Among a lot of possibilities of group decomposition, we assume that $SO(32)$ gauge group decomposes as follows:
\begin{align}
SO(32)\rightarrow SO(2m)\times SO(32-2m)\rightarrow SU(3)_C\times SU(2)_L\times \Pi_{a=1}^{m-3}U(1)_a\times SO(32-2m), 
\label{eq:group1}
\end{align}
where $SO(2m)$ gauge group includes the SM one. 
Now $U(1)$s descend from $U(m)\subset SO(2m)$ and their number is restricted to be $a=1,2,\cdots m-3$, where $3$ is the rank of non-abelian gauge groups in the SM. 
In particular we focus on line bundles $L_a$ each with structure group $U(1)_a$, that is 
the internal bundle of the form
\begin{align}
W=\oplus_a L_a.
\end{align}
Then, $U(1)_a$ gauge fluxes are inserted into the Cartan direction of $SO(2m)$ to realize the SM-like 
gauge group. We expand $U(1)_a$ gauge field strengths ${\rm tr}(\bar{F}_a)$ in the basis of K$\mathrm{\ddot{a}}$hler form $w_i$, $i=1,2,\cdots, h^{1,1}$ 
with $h^{1,1}$ being a hodge number of CY, namely 
\begin{align}
{\rm tr}(\bar{F}_a)=2\pi \sum_{i=1}^{h^{1,1}}{\rm tr}(T_a)m_a^{(i)}w_i,
\end{align}
where $T_a$ are $U(1)_a$ generators descending from $U(m)\subset SO(2m)$. 
Here and in what follows, ``tr'' denotes the trace in the fundamental representation and 
$m_a^{(i)}$ are integers constrained by the Dirac quantization condition.  
Note that the hypercharge is a linear combination of $U(1)_a$, namely
\begin{align}
U(1)_Y=\sum_a f_a U(1)_a,
\end{align}
with normalization factors $f_a$. 
Throughout this paper, we assume the uncorrelated $U(1)_a$ line bundles, otherwise the $U(1)$ gauge groups are enhanced to 
be a non-abelian one.

According to the group decomposition~(\ref{eq:group1}), the adjoint representation of $SO(32)$, corresponding to the massless mode 
in the heterotic string, decomposes under $SO(2m)\times SO(32-2m)$,
\begin{align}
496\rightarrow ({\rm Adj}_{SO(2m)}, 1)\oplus (2m, 32-2m)\oplus (1, {\rm Adj}_{SO(32-2m)}),
\label{eq:496general}
\end{align}
where the adjoint representation of $SO(2m)$, ${\rm Adj}_{SO(2m)}$, is expected to include the SM particles, whereas 
the vector and adjoint representations of $SO(32-2m)$ correspond to the exotic particles. 
On this line bundle background, the net number of chiral zero-modes with $U(1)_a$ charges $Y_a$ is counted by the index
\begin{align}
%\chi (\otimes_{a=1}^n L_a^{Y_a}) =\frac{1}{6}d_{ijk}\left( \sum_{a=1}^n Y_a m_a^{(i)}\right)\left( \sum_{b=1}^n Y_b m_b^{(i)}\right)
%\left( \sum_{c=1}^n Y_c m_c^{(i)}\right)+\sum_{a=1}^n \frac{1}{12}c_{2i}(T{\cal M})Y_am_a^{(i)}. 
\chi (\otimes_{a=1}^n L_a^{Y_a}) =\frac{1}{6}\sum_{a,b,c}X_{abc}Y_aY_bY_c + \frac{1}{12}\sum_a Z_aY_a, 
\label{eq:index}
\end{align}
where we consider internal bundles $\otimes_{a=1}^n L_a^{Y_a}$ and for the later purpose, we define
\begin{align}
Z_a \equiv \sum_{i=1}^{h^{1,1}}c_{2,i}m_a^{(i)},\qquad X_{abc}\equiv \sum_{i,j,k=1}^{h^{1,1}}d_{ijk}m_a^{(i)}m_b^{(j)}m_c^{(k)}.
\end{align}
Here, $d_{ijk}$ are the intersection numbers of the basis of two-forms $w_i$ and the second Chern class of the tangent bundle of CY 
threefolds is expanded in their Hodge dual four-forms $\hat{w}^i$, namely $c_2(T{\cal M})=\sum_i c_{2,i}\hat{w}^i$. 
Variables \{$X_{abc}$, $Z_a$\} are written in terms of the internal gauge fluxes $m_a^{(i)}$ along $U(1)_a$ with generators $T_a$ descending from $U(m)\subset SO(2m)$. 
Since $X_{abc}$ are totally symmetric tensors with respect to $a,b,c$ from the fact that $d_{ijk}$ are totally symmetric 
tensors with respect to $i,j,k$, 
we note that variables \{$X_{abc}$, $Z_a$\} consist of $_{m-3} C _3$+$2(_{m-3} C _2)$+$_{m-3} C _1$=$_{m-1} C _3$ 
and $m-3$ degrees of freedom, totally $_{m-1} C _3 +m-3$, determined by the values of gauge fluxes and topological data of CY. 

The aim of this work is to search for variables \{$X_{abc}$, $Z_a$\} leading to the three-generation SM without specifying 
the topological data of CY. 
An advantage of this approach is the possibility to perform the systematic search on general CY manifolds. 
%However, the internal bundles should satisfy several consistency conditions of heterotic string enumerated as follows. 
Before going to the detailed analysis, we remark three consistency conditions in the four-dimensional 
effective action of heterotic string. (For more details, see, e.g., Refs.~\cite{Blumenhagen:2005ga,Blumenhagen:2005pm}.) 
First one is the ``K-theory condition'' to admit the spinorial representation in the first excited mode~\cite{Witten:1998cd,Uranga:2000xp}
\begin{align}
c_1(W)=\sum_a {\rm tr}(T_a)m_a^{(i)}w_i=2\kappa^{(i)}w_i\in H^2({\rm CY}, 2\mathbb{Z}),
\label{eq:Kth}
\end{align}
%where $c_1(L_a)^{(i)}$ represents the first Chern number of $U(1)$ line bundles $L_a$ on the basis of K$\mathrm{\ddot{a}}$hler form, 
where $c_1(W)$ is the first Chern class of the total internal bundle $W$ and $\kappa^{(i)}$ denote integers.

%The last consistency condition is the stability condition of our discussing four-dimensional effective action. 
Second one is the stability condition of our discussing four-dimensional effective action. 
Stability of the effective action requires a positive number of heterotic five-branes, constraining 
the background curvatures through the anomaly cancellation condition,
\begin{align}
{\rm ch}_2(W)+c_2(T{\cal M})= \sum_i N_i \hat{w}^i, 
\label{eq:tad}
\end{align}
where ${\rm ch}_2(W)$ is the second Chern character of the internal bundle 
and $N_i$ is the number of heterotic five-branes wrapping the internal holomorphic two-cycles on the CY threefold. 
Note that, in the perturbative heterotic string vacua ($N_i=0$), by multiplying the stability condition~(\ref{eq:tad}) by $m_a^{(i)}$, 
we obtain
\begin{align}
\sum_{b}\frac{1}{2}{\rm tr}(T_b)^2X_{abb}+Z_a=0
\label{eq:tadred}
\end{align}
for all $a=1,2,\cdots, m-3$. 

In addition, internal bundles have to satisfy the zero-slope poly-stability conditions, namely $D$-term conditions associated 
to $U(1)_a$ gauge symmetries,
\begin{align}
%D_a %&\propto \frac{1}{2}\int_{\rm CY} J\wedge J\wedge {\rm tr}(\bar{F}_a)+\frac{e^{2\phi_{10}}l_s^4}{6(2\pi)^2}\int_{\rm CY}
%\left( {\rm tr}(\bar{F}_a^3)-\frac{1}{16}{\rm tr}(\bar{F}_a)\wedge {\rm tr}(\bar{R}^2)\right)
%\nonumber\\
D_a  &\propto \frac{1}{2}\sum_{i,j,k}d_{ijk}t_it_j{\rm tr}(T_a)m_a^{(k)}+\frac{e^{2\phi_{10}}}{6}
\left( \sum_{i,j,k}{\rm tr}(T_a^3)d_{ijk}m_a^{(i)}m_a^{(j)}m_a^{(k)}+\frac{1}{8}\sum_i{\rm tr}(T_a)m_a^{(i)}c_{2,i}\right)
=0,
\label{eq:Dterm}
\end{align}
where $\phi_{10}$ is the ten-dimensional dilaton and the K$\mathrm{\ddot{a}}$hler form is now expanded as 
$J=l_s^2 \sum_i t_i w_i$ with $t_i$ being the K$\mathrm{\ddot{a}}$hler moduli in string units 
$l_s=2\pi \sqrt{\alpha^\prime}=1$.\footnote{Note that we now use a different notation for the K$\mathrm{\ddot{a}}$hler moduli, 
$\mathcal{T}_i=\frac{1}{2\pi}(t_i+ib_i^{(0)})$ compared with Ref.~\cite{Blumenhagen:2005pm}. Here, $b_i^{(0)}$ denote the model-dependent axions.} 
Here, we include the one-loop correction to $D$-terms~\cite{Blumenhagen:2005ga,Blumenhagen:2005pm}.

%Second one is the hypercharge masslessness conditions. 
The last one is the hypercharge masslessness conditions. 
On the non-trivial gauge background, the internal gauge fluxes induce the St$\mathrm{\ddot{u}}$ckelberg couplings between string axions 
and the hypercharge gauge boson through the Green-Schwarz terms~\cite{Green:1984bx,Ibanez:1986xy}. 
It is known that some of $U(1)$s are anomalous due to the internal gauge fluxes 
and their number is counted by the rank of $U(1)$ mass matrix in string units 
$l_s=2\pi\sqrt{\alpha'}=1$~\cite{Blumenhagen:2005ga,Blumenhagen:2005pm},
\begin{align}
M_{ai}=\left\{
\begin{array}{c}
2\pi\,{\rm tr}(T_a^2)m_a^{(i)}\qquad {\rm for}\,i=1,2,\cdots, h^{1,1}
\\
\sum_{b,c,d}\frac{1}{6}{\rm tr}(T_aT_bT_cT_d)X_{bcd}+\frac{1}{24}{\rm tr}(T_a^2)Z_a
\qquad {\rm for}\,i=0
\end{array}
\right.
,
\label{eq:Mai}
\end{align}
where the first and second lines are coming from St$\mathrm{\ddot{u}}$ckelberg couplings of K$\mathrm{\ddot{a}}$hler axions and dilaton axion, respectively.
To ensure the masslessness of the hypercharge direction $U(1)_Y=\sum_a f_a U(1)_a$, we impose two constraints originating from 
K$\mathrm{\ddot{a}}$hler axions and dilaton axion,
\begin{align}
&\sum_{a} {\rm tr}(T_a^2) f_a m_a^{(i)}=0, 
\label{eq:hyp1}
\\
&\sum_{a,b,c,d} {\rm tr}(T_aT_bT_cT_d)f_a X_{bcd}=0,
\label{eq:hyp2}
\end{align}
where we note that $\sum_{a} {\rm tr}(T_a^2) f_a Z_a=0$ is satisfied under the constraint $(\ref{eq:hyp1})$. 

Let us take a closer look at the K-theory condition~(\ref{eq:Kth}) and hypercharge masslessness conditions~(\ref{eq:hyp1}) 
which are summarized as
\begin{align}
\begin{pmatrix}
{\rm tr}(T_1) & {\rm tr}(T_2)\\
{\rm tr}(T_1^2)f_1 & {\rm tr}(T_2^2)f_2 \\
\end{pmatrix}
\begin{pmatrix}
m_1^{(i)}\\
m_2^{(i)}\\
\end{pmatrix}
=-
\begin{pmatrix}
\sum_{A=3}^{m-3}{\rm tr}(T_A)m_A^{(i)}\\
\sum_{A=3}^{m-3}{\rm tr}(T_A^2)m_A^{(i)}f_A\\
\end{pmatrix}
+
\begin{pmatrix}
2\kappa^{(i)}\\
0\\
\end{pmatrix}
.
\end{align}
Thus two $U(1)$ magnetic fluxes $m_{\alpha}^{(i)}$ ($\alpha=1,2$) can be determined by other $U(1)$ fluxes
\begin{align}
m_\alpha^{(i)}=\sum_{A=3}^{m-3} K_{\alpha A}m_A^{(i)} +V_{\alpha}\kappa^{(i)},
\label{eq:malphai}
\end{align}
where 
\begin{align}
K_{1 A}&=\frac{{\rm tr}(T_2){\rm tr}(T_A^2)f_A-{\rm tr}(T_2^2){\rm tr}(T_A)f_2}
{{\rm tr}(T_1){\rm tr}(T_2^2)f_2-{\rm tr}(T_2){\rm tr}(T_1^2)f_1}
,\qquad 
K_{2 A}=\frac{{\rm tr}(T_1^2){\rm tr}(T_A)f_1-{\rm tr}(T_1){\rm tr}(T_A^2)f_A}
{{\rm tr}(T_1){\rm tr}(T_2^2)f_2-{\rm tr}(T_2){\rm tr}(T_1^2)f_1}
,\nonumber\\
V_{1}&=\frac{2{\rm tr}(T_2^2)f_2}
{{\rm tr}(T_1){\rm tr}(T_2^2)f_2-{\rm tr}(T_2){\rm tr}(T_1^2)f_1},
\qquad
V_{2}=-\frac{2{\rm tr}(T_1^2)f_1}
{{\rm tr}(T_1){\rm tr}(T_2^2)f_2-{\rm tr}(T_2){\rm tr}(T_1^2)f_1}.
\end{align}
Here and in what follows, two generators $T_{1,2}$ are chosen such that ${\rm tr}(T_1){\rm tr}(T_2^2)f_2-{\rm tr}(T_2){\rm tr}(T_1^2)$
$f_1\neq 0$. 
It then allows us to rewrite variables $\{X_{\alpha BC}, X_{\alpha \beta C}, X_{\alpha \beta \gamma}, Z_\alpha\}$ in terms of others, namely 
\begin{align}
%X_{\alpha BC}=d_{ijk}m_\alpha^{(i)}m_B^{(j)}m_C^{(k)}=K_{\alpha A}X_{ABC}+V_\alpha X^\prime_{BC},\qquad
%d_{ijk}m_\alpha^{(i)}m_\beta^{(j)}m_C^{(k)}=K_{\beta B}X_{\alpha BC}
&X_{\alpha BC}=\sum_A K_{\alpha A}X_{ABC}+V_\alpha X^\prime_{BC},\qquad
X_{\alpha \beta C}=\sum_A K_{\alpha A}X_{\beta AC}+V_\alpha\left(\sum_B K_{\beta B}X^\prime_{BC}+V_\beta X^{''}_C\right),\nonumber\\
&X_{\alpha \beta\gamma}=\sum_C K_{\gamma C}X_{\alpha \beta C}+K_{\alpha A}V_\gamma \left(\sum_B K_{\beta B}X^\prime_{BA}+V_\beta X^{''}_A\right)
+V_\alpha V_\gamma \left(\sum_{B}K_{\beta B}X_B^{''}+V_\beta X^{'''}\right),
\nonumber\\
&Z_\alpha=\sum_A K_{\alpha A} Z_A+V_\alpha Z^\prime,
\label{eq:XZconst}
\end{align}
where 
\begin{align}
&X^\prime_{AB}=\sum_{i,j,k}d_{ijk}\kappa^{(i)}m_A^{(j)}m_B^{(k)},\qquad
X^{''}_A=\sum_{i,j,k} d_{ijk}\kappa^{(i)}\kappa^{(j)}m_A^{(k)},\qquad
X^{'''}=\sum_{i,j,k} d_{ijk}\kappa^{(i)}\kappa^{(j)}\kappa^{(k)},
\nonumber\\
&Z^\prime=\sum_{i} c_{2,i}\kappa^{(i)}.
\end{align}

As a result, independent variables are $\{X_{ABC}, X^\prime_{AB}, X^{''}_A, X^{'''}, Z_A, Z^\prime\}$ not $\{X_{abc}, Z_a\}$. 
Thus, the number of variables reduces to $m(m-4)(m-5)/6+2(m-4)$ from $_{m-1} C _3 +m-3$. 
We now also use the totally symmetric properties of $d_{ijk}$. 

In addition to the theoretical requirements such as K-theory condition~(\ref{eq:Kth}) and hypercharge masslessness 
conditions~(\ref{eq:hyp1}), variables $\{X_{ABC}, X^\prime_{AB}, X^{''}_A, X^{'''}, Z_A, Z^\prime\}$ are further constrained by phenomenological 
requirements. 
To realize phenomenologically consistent models, we impose the three generations of quarks and charged leptons\footnote{At 
this stage, we have not distinguished between the Higgsino fields and the charged leptons. } and 
no chiral exotics, namely
\begin{align}
\chi_Q=\chi_L=\chi_{u^c}=\chi_{d^c}=\chi_{e^c}=-3|\Gamma|, \qquad  \chi_{\rm exotic}^{(p)}=0,
\label{eq:indcond}
\end{align}
for all $1\leq p\leq p_{\rm ex}$, where the number of chiral exotics $p_{\rm ex}$ depends on the 
branching of $SO(32)$ but is at least $p_{\rm ex}\geq m-3$ appearing from 
exotic states $(2m, 32-2m)$ in Eq.~(\ref{eq:496general}). 
We have taken into account the order of freely-acting discrete symmetry group of CY threefolds $|\Gamma|$ 
in order to be applicable to the model building on non-simply-connected CY threefolds. 
The above phenomenological requirements constrain 
variables $\{X_{ABC}, X^\prime_{AB}, X^{''}_A, X^{'''}, Z_A, Z^\prime\}$ through Eq.~(\ref{eq:index}). 
It turns out that in the case with $m=6$, i.e., $SO(32)\rightarrow SO(12)\times SO(20)\rightarrow 
SU(3)_C\times SU(2)_L\times \Pi_{a=1}^3 U(1)_a \times SO(20)$, 
total six variables are not enough to satisfy at least eight index conditions~(\ref{eq:indcond}) and 
remaining hypercharge masslessness condition~(\ref{eq:hyp2}) in general. 
Here, we have not considered the other stability conditions~(\ref{eq:tad}) and (\ref{eq:Dterm}) which depend 
on the number of five-branes and values of moduli fields. 

To simplify our analysis, we focus on $m=8$ case, namely $SO(32)\rightarrow SO(16)\times SO(16)^\prime\rightarrow 
SU(3)_C\times SU(2)_L\times \Pi_{a=1}^5 U(1)_a \times SO(16)^\prime$, including the $m=7$ case. 
Note that such a visible $SO(16)$ gauge group can be also 
embedded into the T-dual $E_8\times E_8$ and non-supersymmetric $SO(16)\times SO(16)$ heterotic string theories, 
taking into account the spinor representation of $SO(16)$ as a massless mode. 
The following analysis is then applicable to other heterotic string theories.

%%%%%%%%%%%%%%%%%%%%%%%%%%%%%%%%%%%%%%%%%%%%%%%%%%%%%%%%%%%%%%%%%%%%%%%%%%%%%%%%%%%%%%%%%%%%%%%%%%%%%%%%%%%%%%
\subsection{Embedding the SM gauge group into $SO(16)$}
\label{subsec:2.2}

Then, let us consider the detailed group decomposition of $SO(32)$ using the internal gauge fluxes. 
As a concrete example, we directly embed the SM gauge group into $SO(16) \subset SO(32)$ as mentioned before. 
Among the branching of $SO(16)$, for illustrative purpose, we focus on the following decomposition of $SO(32)$\footnote{For other 
gauge branchings, see, Appendix~\ref{app}.}: 
\begin{align}
SO(32)\rightarrow SO(16)\times SO(16)' \rightarrow SU(3)_C \times SU(2)_L \times \Pi_aU(1)_a  \times SO(16)',
\label{eq:T32111_1}
\end{align}
where the $U(1)_a$ directions are chosen such that
\begin{align}
U(1)_1&: (1,1,1,0,0,0,0,0;0,\cdots,0),
\nonumber\\
U(1)_2&: (0,0,0,1,1,0,0,0;0,\cdots,0),
\nonumber\\
U(1)_3&: (0,0,0,0,0,1,0,0;0,\cdots,0),
\nonumber\\
U(1)_4&: (0,0,0,0,0,0,1,0;0,\cdots,0),
\nonumber\\
U(1)_5&: (0,0,0,0,0,0,0,1;0,\cdots,0),
\label{eq:T32111}
\end{align}
in the basis of Cartan directions of $SO(32)$ $H_i (i=1,2,\cdots, 16)$. The Cartan directions of $SU(3)_C$ and $SU(2)_L$ are respectively 
taken as $H_1-H_2$, $H_2-H_3$ and $H_4-H_5$ in our analysis and $SO(32)$ roots correspond to $(\underline{\pm 1, \pm 1, 0,\cdots, 0})$ under $H_i$, where the underline denotes the possible permutation. 
Under the decomposition $SO(32)\rightarrow SO(16)\times SO(16)'$, the adjoint representation of $SO(32)$ $496$ decomposes as follows:
%Since $SO(32)$ roots correspond to $(\underline{\pm 1, \pm 1, 0,\cdots, 0})$ under $H_i$, 
%the adjoint representation of SO(32) decomposes as follows:
\begin{align}
496\rightarrow (120,1)\oplus (16, 16')\oplus (1, 120'),
\end{align}
where the adjoint representation of $SO(16)$ includes the visible sector and others correspond to the hidden sector. 
In this decomposition, we find that there exist eight options to take the correct hypercharge, 
but for our purpose, we take the specific hypercharge direction\footnote{It is possible to consider other hypercharge directions: 
$U(1)_Y=-\frac{1}{6}U(1)_1 \pm \frac{1}{2}\left(U(1)_3 \pm U(1)_4 \pm U(1)_5\right)$, where the sign is taken in the random order. 
The algorithm to determine the hypercharge direction is discussed in Appendix~\ref{app}.},
\begin{align}
U(1)_Y=-\frac{1}{6}U(1)_1 -\frac{1}{2}\biggl(U(1)_3 -U(1)_4 +U(1)_5\biggl),
\label{eq:U(1)Y32111}
\end{align}
under which the spectrum consists of the visible sector:
\begin{align}
Q :~& (3, 2)_{-1,1,0,0,0} \oplus (3, 2)_{-1,-1,0,0,0}, \nonumber\\
L :~&(1, 2)_{0,\pm 1,1,0,0} \oplus (1, 2)_{0,\pm 1, 0, -1,0} \oplus (1, 2)_{0,\pm 1,0, 0, 1},\nonumber\\
u^c :~&(\bar{3}, 1)_{1,0, 1,0,0} \oplus  (\bar{3}, 1)_{1,0, 0, -1,0} \oplus (\bar{3}, 1)_{1,0, 0,0, 1}, \nonumber\\
d^c :~&(\bar{3}, 1)_{1,0, -1,0,0} \oplus (\bar{3}, 1)_{1,0, 0, 1,0} \oplus (\bar{3}, 1)_{1,0, 0,0, -1},\nonumber\\
e^c:~&(1, 1)_{0,0,-1, 1,0} \oplus (1, 1)_{0,0,-1,0,-1} \oplus (1, 1)_{0,0,0, 1, -1},\nonumber\\
n^c:~&(1, 1)_{0,-2,0,0,0} \oplus (1, 1)_{0,0,-1,-1,0} \oplus (1, 1)_{0,0, -1,0,1}\oplus (1, 1)_{0,0, 0,1,1},\nonumber\\
{\rm Exotic\:state}:~&(\bar{3}, 1)_{-2,0, 0,0,0},
\end{align}
and the hidden sector:
\begin{align}
&(3, 1,16')_{1,0,0,0,0} \oplus (1, 2,16')_{0,1,0,0,0}  \oplus (1, 1,16')_{0,0,1,0,0} \oplus \nonumber\\
&(1, 1,16')_{0,0,0,1,0} \oplus (1, 1,16')_{0,0,0,0,1} \oplus (1, 1,120')_{0,0, 0,0,0}.
\end{align}
Now the subscript indices denote the $U(1)_a$ charges $Y_a$ and the particle $(\bar{3}, 1)_{-2,0, 0,0,0}$ 
is an exotic particle due to the absence of the Yukawa couplings as indicated by their weights~(\ref{eq:weight}) in Appendix~\ref{app}. 
Note that we cannot distinguish the particle and anti-particles for SM singlets $n^c$ at this stage. 
We thus show one of the concrete examples out of eight possibilities.

To obtain the SM-like spectra without chiral exotics, 
we require 
\begin{align}
Q :~& \chi (L_1^{-1}\times L_2) + \chi (L_1^{-1}\times L_2^{-1})=-3|\Gamma|, \nonumber\\
L :~& \sum_{s=\pm}\chi (L_2^{s}\times L_3^{-1})+ \sum_{s=\pm}\chi (L_2^{s}\times L_4^{1})
+\sum_{s=\pm}\chi (L_2^{s}\times L_5^{1})=-3|\Gamma|, \nonumber\\
u^c :~& \chi (L_1\times L_3) + \chi (L_1\times L_4^{-1}) + \chi (L_1\times L_5)=-3|\Gamma|, \nonumber\\
d^c :~& \chi (L_1\times L_3^{-1}) + \chi (L_1\times L_4) + \chi (L_1\times L_5^{-1})=-3|\Gamma|, \nonumber\\
e^c:~& \chi (L_3^{-1}\times L_4)+ \chi (L_3^{-1}\times L_5^{-1}) + \chi (L_4\times L_5^{-1})=-3|\Gamma|,
\label{eq:indvis}
\end{align}
and 
\begin{align}
\chi (L_1)=\chi (L_2)=\chi (L_3)=\chi (L_4)=\chi (L_5)=\chi (L_1^{-2})=0,
\label{eq:indhid}
\end{align}
employing the index theorem on CY threefolds. 
%Then we further impose
%\begin{align}
%\chi (L_1\times L_3)=\chi (L_1^{-2})=0,
%\label{eq:indhid2}
%\end{align}
%which also holds for other branchings of $SO(16)$ as confirmed in the latter analysis. 
The above index formulae are applicable to both simply-and non-simply-connected CY threefolds with 
the order of freely-acting discrete symmetry group $|\Gamma|$ 
and here we have not counted the number of Higgs/Higgsino fields.

Although we have focused on the particular branching of $SO(32)$ with specific hypercharge direction, 
we systematically investigate several gauge embeddings with possible hypercharge directions 
listed in Table~\ref{tab:group} of Appendix~\ref{app}. 
In the next section~\ref{subsec:2.3}, we solve the eleven index conditions~(\ref{eq:indvis}), (\ref{eq:indhid}) 
and remaining hypercharge masslessness condition~(\ref{eq:hyp2}) in terms of $24$ variables $\{X_{ABC}, X^\prime_{AB}, X^{''}_A, X^{'''}, Z_A, Z^\prime\}$ against several branchings of $SO(32)$.

%%%%%%%%%%%%%%%%%%%%%%%%%%%%%%%%%%%%%%%%%%%%%%%%%%%%%%%%%%%%%%%%%%%%%%%%%%%%%%%%%%%%%%%%%%%%%%%%%%%%%%%%%%%%
%%%%%%%%%%%%%%%%%%%%%%%%%%%%%%%%%%%%%%%%%%%%%%%%%%%%%%%%%%%%%%%%%%%%%%%%%%%%%%%%%%%%%%%%%%%%%%%%%%%%%%%%%%%%
\subsection{General formula}
\label{subsec:2.3}

We are now ready to search for 24 variables $\{X_{ABC}, X^\prime_{AB}, X^{''}_A, X^{'''}, Z_A, Z^\prime\}$ with $A=3,4,5$, satisfying 
the eleven index conditions~(\ref{eq:indvis}), (\ref{eq:indhid}) and the remaining hypercharge masslessness condition~(\ref{eq:hyp2}). 
Since the stability conditions depend on the detail of CY data, for the time being, we focus on only 
K-theory condition and hypercharge masslessness conditions. 
We remark that the stability conditions have to be checked for each CY threefold. 
%An advantage of this approach is the possibility to perform the systematic search on general CY manifolds. 
The details of the algorithm computing the particle spectrum is shown in Appendix~\ref{app}. 

On simply-connected CY threefolds, our systematic search reveals that a direct flux breaking scenario is available for 
the following decompositions:
\begin{align}
SO(16)&\rightarrow 
\left\{
 \begin{array}{l}
 SO(6)\times SO(4)\times SO(2)^3
\nonumber\\
 SO(10)\times SO(6)\rightarrow SU(5)\times SU(3)\times U(1)^2
\nonumber\\
SO(8)\times SO(4)\times SO(4)\rightarrow SU(4)_C\times SU(2)_L\times SU(2)\times U(1)^3
\nonumber\\
SU(4)_C\times SU(4)\times U(1)^2\rightarrow SU(4)_C\times SU(2)_L\times SU(2)\times U(1)^3
\nonumber\\
SO(6)\times SO(6)\times SO(4)\rightarrow SU(3)_C\times SU(3)\times SU(2)\times U(1)^3
\nonumber\\
SO(8)\times SO(6)\times SO(2)\rightarrow SU(4)_C\times SU(3)\times U(1)^3
\nonumber\\
SO(8)\times SO(4)\times SO(2)^2\rightarrow SU(4)_C\times SU(2)_L\times U(1)^3
\nonumber\\
SO(6)\times SO(10)\rightarrow SU(3)_C\times SU(5)\times U(1)^2
\nonumber\\
SO(6)\times SO(4)\times SO(4)\times SO(2)\rightarrow SU(3)_C\times SU(2)_L\times SU(2)\times U(1)^3
\nonumber\\
SO(6)\times SO(6)\times SO(2)^2\rightarrow SU(3)_C\times SU(3)\times U(1)^4
\end{array}
\right. 
\nonumber\\
&\rightarrow G_{\rm SM}\times U(1)^4,
\label{eq:SO(16)2}
\end{align}
with $G_{\rm SM}=SU(3)_C\times SU(2)_L\times U(1)_Y$. 
In the case with the first branch of Eq.~(\ref{eq:SO(16)2}) corresponding to the model of Sec.~\ref{subsec:2.2}, 
the following 24 variables satisfy all the requirements:
\begin{align}
&X_{333}=p_1, X_{334}=p_2, X_{335}=p_3, X_{344}=p_4, X_{345}=3, X_{355}=p_5, X_{444}=p_6,
\nonumber
\\
&X_{445}=p_7, X_{455}=-6-p_2+p_3+p_4+p_5+p_7, X_{555}=-p_1+p_6, X^\prime_{33}=p_8, X^\prime_{34}=p_{9}, X^\prime_{35}=p_{10},  
\nonumber
\\
&X^\prime_{44}=p_{11},  X^\prime_{45}=p_{12}, X^\prime_{55}=p_{13}, X^{''}_{3}=p_{14},X^{''}_{4}=p_{15}, 
\nonumber
\\
&X^{''}_{5}=-3+2p_3-3p_4+2p_5+3p_6-3p_7-2p_8 +6p_9-4p_{10}-4p_{11} +6p_{12}-2p_{13}-p_{14}+p_{15}, 
\nonumber
\\
&X^{'''}=p_{16}, Z_3=-2p_1,  Z_4=-2p_6, Z_5=2p_1-2p_6, 
\nonumber
 \\
 &Z^{'}=18-6p_3+6p_4-6p_5-6p_6+6p_7+6p_8-12p_9+12p_{10}-12p_{12}+6p_{13}+6p_{15}-2p_{16},
 \label{eq:XZ32111}
\end{align}
%\begin{align}
%&X_{333}=p_1, X_{334}=p_2, X_{335}=p_3, X_{344}=p_4, X_{345}=p_5, X_{355}=1+11p_1+5p_4+16p_6, X_{444}=2p_1+3p_6,
%\nonumber
%\\
%&X_{445}=p_7, X_{455}=-1+7p_1+4p_4+11p_6, X_{555}=p_8, X^\prime_{33}=p_9, X^\prime_{34}=p_{10}, X^\prime_{35}=p_{11},  
%X^\prime_{44}=p_{12}, 
%\nonumber
%\\
%&X^\prime_{55}=-3+408p_{1}+54p_2+192p_{4}+600p_6-3p_9-12p_{10}-27p_{12}, 
%\nonumber
%\\
%&X^{''}_{3}=-1+43p_1-48p_2-20p_4+80p_6+6p_9+12p_{10},
%\nonumber
%\\
%&X^{''}_{4}=1-85p_1-6p_2-40p_4-128p_6+6p_{10}+12p_{12}, 
%X^{''}_{5}=-6p_3-36p_5-57p_7+p_8 +6p_{11} +12p_{13}, 
%\nonumber
%\\
%&X^{'''}=9-684p_1 -378p_2-576p_4-936p_6 +27p_9+ 108p_{10} +108p_{12},
%\nonumber
%\\
%&Z_3=-66p_1 -48p_4 -96p_6, 
% Z_4=-48p_1 -6p_2 -12p_4 -72p_6, 
% Z_5=-6p_3 -12p_5 -6p_7 -2p_8, 
%\nonumber
% \\
% &Z^{'}=-486p_1 -36p_2 -216p_4 -720p_6,
%\end{align}
where $p_{1},\cdots, p_{16}$ are integers and we now consider the specific hypercharge direction~(\ref{eq:U(1)Y32111}). 
For other gauge branchings, see, Table~\ref{tab:hflux} of Appendix~\ref{app}. 
As mentioned before, some of $U(1)$s are anomalous due to the internal gauge fluxes 
and their number is counted by the rank of $U(1)$ mass matrix~(\ref{eq:Mai}). 
Remarkably, for all the gauge decompositions of Table~\ref{tab:hflux}, the dilaton-axion induced mass term in Eq.~(\ref{eq:Mai}) 
turns out to be 
\begin{align}
M_{a0}=
\sum_{b,c,d}\frac{1}{6}{\rm tr}(T_aT_bT_cT_d)X_{bcd}+\frac{1}{24}{\rm tr}(T_a^2)Z_a =-\frac{1}{24}{\rm tr}(T_a^2)Z_a, 
\end{align}
due to the correlation between $X_{abc}$ and $Z_a$. 

Let us consider in more detail the $\kappa^{(i)}=0$ case which means that the right-handed side of K-theory condition~(\ref{eq:Kth}) vanishes. 
In such a case, independent variables are $\{X_{ABC}, Z_A\}$, since other variables are written by
\begin{align}
%X_{\alpha BC}=d_{ijk}m_\alpha^{(i)}m_B^{(j)}m_C^{(k)}=K_{\alpha A}X_{ABC}+V_\alpha X^\prime_{BC},\qquad
%d_{ijk}m_\alpha^{(i)}m_\beta^{(j)}m_C^{(k)}=K_{\beta B}X_{\alpha BC}
&X_{\alpha BC}=\sum_{A}K_{\alpha A}X_{ABC},\quad
X_{\alpha \beta C}=\sum_{A,B}K_{\alpha A}K_{\beta B}X_{ABC},\quad 
X_{\alpha \beta\gamma}=\sum_{A,B,C}K_{\alpha A}K_{\beta B}K_{\gamma C}X_{ABC},\quad 
\nonumber\\
&Z_\alpha=\sum_{A}K_{\alpha A} Z_A,
\end{align}
from which two $U(1)$ gauge boson mass terms are provided by the other one, i.e., $M_{\alpha i}=\sum_A K_{\alpha A}M_{A i}$. 
Thus, the maximal rank of $U(1)$ gauge boson matrix is $3$. One of the massless $U(1)$ directions will be identified 
with $U(1)_{B-L}$ in addition to $U(1)_Y$. 
On the other hand, in the $\kappa^{(i)}\neq 0$ case, the rank of $U(1)$ gauge boson matrix is generically $4$ when $h^{1,1}\geq 3$ 
and the remaining gauge symmetry consists of $SU(3)_C\times SU(2)_L\times U(1)_Y \times SO(16)^\prime$. 
%In the next section, we explore three-generation models with an emphasis on $\kappa^{(i)}\neq 0$ to 
%avoid the existence of massless $U(1)$s except for $U(1)_Y$. 

Finally, we comment on the direct flux breaking scenario on non-simply-connected CY threefolds. 
In contrast to the simply-connected CY cases, some special non-simply-connected CY 
threefolds, specifically $|\Gamma|=5\mathbb{Z}_{>0}$, allow intermediate GUT-like models without chiral exotics.  
The detail of phenomenologically acceptable CYs is shown in Table~\ref{tab:group} of Appendix~\ref{app}.

\section{Phenomenological aspects of three-generation models}
\label{sec:3}

The obtained general formula in Sec.~\ref{subsec:2.3} is applicable to general CY threefolds. 
In this section, we discuss three-generation models on a specific CICY.

\subsection{Three-generation models on simply-connected CY threefolds}
\label{subsec:3.1}
We consider simply-connected CY threefolds, namely $|\Gamma|=1$. 
%As discussed in Sec.~\ref{subsec:2.3}, the $SU(5)$ and $SO(10)$ GUT-like spectrum cannot be achieved on this setup 
%and 
Among known $7890$ CICYs, we consider the following CICY called $\#$ 7247 in the list of \cite{Oxford},
\begin{align}
\begin{matrix}
\mathbb{P}^2\\
\mathbb{P}^2\\
\mathbb{P}^2\\
\mathbb{P}^2\\
\end{matrix}
\!\left[
\begin{matrix}
1 & 1 & 1 & 0 & 0 \\
1 & 1 & 0 & 1 & 0 \\
1 & 1 & 0 & 0 & 1 \\
0 & 0 & 1 & 1 & 1 \\
\end{matrix}
\right]
^{4, 40}_{-72},
\label{eq:7247}
\end{align} 
where the above configuration matrix characterizes how to embed the CY manifold in 
four projective spaces $\mathbb{P}^2$. 
The superscript and subscript indices denote the hodge number of CY $(h^{1,1},h^{2,1})=(4,40)$ 
and the Euler number of CY $-72$, respectively. (See for details of CICYs, e.g., Ref.~\cite{Hubsch:1992nu}.) 
The topological data of the above CICY is calculated as
\begin{align}
&d_{123}=6,
\qquad
d_{124}=d_{134}=d_{234}=5,
\qquad
d_{112}=d_{113}=d_{122}=d_{133}=d_{223}=d_{233}=3,
\nonumber\\
&d_{114}=d_{144}=d_{224}=d_{244}=d_{334}=d_{344}=2,
\qquad
d_{111}=d_{222}=d_{333}=d_{444}=0,
\end{align}
in the basis of two-forms $w_i$ and 
\begin{align}
c_2(T{\cal M})=(36, 36, 36, 36),
\end{align}
in the basis of dual four-forms $\hat{w}^i$, respectively. 
Note that this CY has $Z_3$ and $Z_3\times Z_3$ discrete symmetries, but we concentrate on the case 
with $|\Gamma|=1$ in what follows. 

For concreteness, we study compactifications of the $SO(32)$ heterotic string on 
the above CICY with line bundles leading to the group decomposition~(\ref{eq:T32111_1}). 
Although the model-building approach is classified into two cases: $\kappa^{(i)}\neq 0$ and $\kappa^{(i)}=0$ appearing in 
the K-theory condition~(\ref{eq:Kth}), in the following analysis, we focus on the case with $\kappa^{(i)}\neq 0$, 
providing just the SM gauge group in the visible sector.

%%%%%%%%%%%%%%%%%%%%%%%%%%%%%%%%%%%%%%%%%%%%%%%%%%%%%%%%%%%%%%%%%%%%%%%%%%%%%%%%%%%%%%%%%%%%%%%%%%%%%%%%%
We search for the internal $U(1)$ gauge fluxes within the range $-1\leq m_A^{(i)}\leq 1$ and $-1\leq \kappa^{(i)} \leq 1$ 
where $i=1,2,3,4$ and $A=3,4,5$, 
constrained by Eqs.~(\ref{eq:indvis}) and (\ref{eq:indhid}), the stability conditions~(\ref{eq:tad}) and (\ref{eq:Dterm}) 
and hypercharge masslessness condition~(\ref{eq:hyp2}). 
Note that other $U(1)_{1,2}$ fluxes are determined by Eq.~(\ref{eq:malphai}). 
It turns out that some of the fluxes satisfy all the requirements. 
For example, under the flux ansatz, 
\begin{align}
&m_1=(1,0,0,-1),\qquad m_2=(1,0,-1,3), \qquad m_3=(0,1,0,-1),\nonumber\\
&m_4=(0,0,1,-1), \qquad m_5=(-1,-1,1,1),
\end{align}
the number of five-branes in the basis of $\hat{w}^i$, 
\begin{align}
N_{i}=(6,30,72,6),
\end{align}
is positive and at the same time, the $D$-term conditions~(\ref{eq:Dterm}) are satisfied 
at the physical domain of K$\mathrm{\ddot{a}}$hler moduli and dilaton, namely $t_i, s>1$ in string units,
\begin{align}
t_1=t_2=t_3=\frac{3+2\sqrt{3}}{3}t_4,\quad
s=\frac{2+\sqrt{3}}{2\pi}t_4,
\end{align}
where the ten-dimensional dilaton field in Eq.~(\ref{eq:Dterm}) is now 
written by $s=e^{-2\phi_{10}}{\cal V}/(2\pi)$ with CY volume ${\cal V}=\sum_{i,j,k}d_{ijk}t_it_jt_k/6$ 
in string units. 
The above fluxes result in the following chiral zero-modes: 
\begin{align}
&Q:~3(3, 2)_{-1,1,0,0,0},\nonumber\\
&L:~3(1, 2)_{0,1,0,0,1},\nonumber\\
&H_u^1:~6(1, 2)_{0, -1, -1, 0,0},\quad H_u^2:~12(1, 2)_{0, 1, -1, 0,0},\quad H_u^3:~21(1, 2)_{0, 1, 0, 0, -1},\nonumber\\
&H_d^1:~15(1, 2)_{0,-1,0, -1, 0},\quad H_d^2:~24(1, 2)_{0, 1, 0, -1, 0},\nonumber\\
&u^c:~3(\bar{3}, 1)_{1,0, 1, 0,0},\nonumber\\
&d^c:~3(\bar{3}, 1)_{1,0, 0, 1, 0},\qquad \nonumber\\
&e^c:~3(1, 1)_{0,0,0,1, -1},
\end{align}
and singlets:
\begin{align}
&n_1^c:~3(1, 1)_{0,0,0,-1,-1},\quad n_2^c:~3(1, 1)_{0, 0, 1, 1,0},\quad n_3^c:~54(1, 1)_{0, 2, 0, 0, 0},
\end{align}
where the particles $n_1^c$ or $n_2^c$ could be identified with right-handed neutrinos. 
Remarkably, we have phenomenologically interesting perturbative Yukawa couplings,
\begin{align}
&QH_u^1 u^c,\quad QH_d^1 d^c,\quad LH_d^1 e^c,\quad LH_u^1 n_1^c n_2^c,
\end{align}
and interestingly, the higher-dimensional proton decay operators are prohibited by the massive $U(1)_{B-L}$ gauge symmetry. 
However, we require the mechanism generating the mass terms for extra Higgs doublets, 
which will be the subject of future work.

%%%%%%%%%%%%%%%%%%%%%%%%%%%%%%%%%%%%%%%%%%%%%%%%%%%%%%%%%%%%%%%%%%%%%%%%%%%%%%%%%%%%%%%%%%%%%%%%%%%%%%%%%%
\subsection{Gauge coupling unification}
\label{subsec:3.2}
So far, we have focused on the number of chiral generations in $SO(32)$ heterotic compactification. 
In this section, we discuss the unification of gauge couplings at the string scale. 
The four-dimensional gauge coupling is extracted from the ten-dimensional $SO(32)$ Yang-Mills action, 
\begin{align}
-\frac{1}{2g_{10}^2}\int e^{-2\phi_{10}}{\rm tr}_{SO(32)}(F\wedge \ast_{10} F),
\end{align}
where $g_{10}^2=4\pi (l_s)^6$ is the ten-dimensional gauge coupling. 
After expanding the ten-dimensional gauge field strength as the four-dimensional part $F$ and internal part $\bar{F}$, 
namely $F\rightarrow F+\bar{F}$, 
the kinetic terms of the four-dimensional gauge bosons become
\begin{align}
-\frac{1}{4}\int s\,{\rm tr}_{SO(32)}(F\wedge \ast_{4} F),
\end{align}
where $s$ is the dilaton field
\begin{align}
s=e^{-2\phi_{10}}{\cal V}/2\pi,
\end{align}
with CY volume ${\cal V}$ in string units.

Since the generator of hypercharge direction listed in all the gauge decompositions of Table~\ref{tab:group} is of the form
\begin{align}
U(1)_Y=\frac{1}{6}{\rm diag}(-1,-1,-1,0,0,\pm 3, \pm 3, \pm 3),
\end{align}
where the sign is taken in the random order, the four-dimensional gauge field strength of the SM gauge group is expanded as 
\begin{align}
F= 
\begin{pmatrix}
\sum_{A=1}^8 F_{SU(3)_C}^A \lambda^A & 0 & 0 & 0 & 0 \\
0 & \sum_{\alpha=1}^3 F_{SU(2)_L}^\alpha \sigma^\alpha & 0 & 0 & 0\\
0 & 0 & 0 & 0 & 0\\
0 & 0 & 0 & 0 & 0\\
0 & 0 & 0 & 0 & 0\\
\end{pmatrix}
+\frac{F_Y}{6}
\begin{pmatrix}
-1_{3\times 3} & 0 & 0 & 0 & 0 \\
0 & 0 & 0 & 0 & 0\\
0 & 0 & \pm 3  & 0 & 0\\
0 & 0 & 0 & \pm 3 & 0\\
0 & 0 & 0 & 0 & \pm 3\\
\end{pmatrix}
,
\end{align}
using the Gell-Mann matrices $\lambda^A$ and Pauli matrices $\sigma^\alpha$. 
In our case, the non-abelian part is normalized as ${\rm tr}_{SO(32)}(T_{SU(3)_C}^AT_{SU(3)_C}^B)=2\delta^{AB}$, 
${\rm tr}_{SO(32)}(T_{SU(2)_L}^\alpha T_{SU(2)_L}^\beta)=2\delta^{\alpha\beta}$ due to 
the fact that the level of Kac-Moody algebra is one. 
The trace of gauge field strength then reads
\begin{align}
{\rm tr}_{SO(32)}(F\wedge \ast_4 F)=2F_{SU(3)_C}^a\wedge \ast_4 F_{SU(3)_C}^a+2F_{SU(2)_L}^m\wedge \ast_4 F_{SU(2)_L}^m+ \frac{5}{3}F_{U(1)_Y}\wedge \ast_4 F_{U(1)_Y}. 
\end{align} 
Note that off-diagonal $U(1)$ gauge couplings are absent in contrast to the $E_8\times E_8$ heterotic string case~\cite{Blumenhagen:2005pm}. 
Taking into account the normalization of $U(1)_Y$ generator, the gauge couplings satisfy
\begin{align}
g_{SU(3)_C}^2=g_{SU(2)_L}^2 =\frac{5}{6}g_{U(1)_Y}^2,
\end{align}
at tree level. 

Thus, we cannot arrive at phenomenologically interesting so-called GUT-relation. 
However, the unification of gauge couplings actually depends on the number of Higgs doublets, radiative corrections to the gauge couplings and supersymmetry-breaking scale. Indeed, the radiative corrections to the non-abelian gauge groups are non-universal in contrast to the 
$E_8\times E_8$ heterotic case~\cite{Blumenhagen:2005ga,Blumenhagen:2005pm}, which allows us to obtain the realistic values of gauge 
couplings at the string scale as demonstrated on toroidal background~\cite{Abe:2015xua}. To evaluate their precise values, we have to discuss the stabilization mechanism of moduli fields and supersymmetry-breaking sector. We will postpone the concrete model building for a future analysis.

%\begin{align}
%Q :~& (0, 0, -1, 1, 0, 0, 0, 0) \oplus (0, 0, -1, 0, -1, 0, 0, 0) \nonumber\\
%L :~& (0, 0, 0, 1, 0, \pm 1, 0, 0) \oplus (0, 0, 0, 1, 0, 0, \pm 1, 0) \oplus (0, 0, 0, 1, 0, 0, 0, \pm 1) \nonumber\\
%u^c,d^c :~& (0, -1, -1, 0, 0,0, 0, 0) \oplus (1, 0, 0, 0, 0, \pm 1, 0, 0) \oplus (1, 0, 0, 0, 0, 0, \pm 1, 0) \oplus (1, 0, 0, 0, 0, 0, 0, \pm 1) %\nonumber\\
%e^c,n^c :~& (0, 0, 0, 1, 1, 0, 0, 0) \oplus (0, 0, 0, 0, 0, \pm 1, 1, 0) \oplus (0, 0, 0, 0, 0, \pm 1, 0, 1) \oplus (0, 0, 0, 0, 0, 0, \pm 1, 1) \nonumber\\
%\end{align}
%under $H_m$ and others correspond to the hidden sector,
%\begin{align}
%16' :~&(1, 0, 0, 0, 0, 0, 0, 0; \underline{\pm 1,0,\cdots,0}) \oplus (0, 0, 0, 1, 0, 0, 0, 0; \underline{\pm 1,0,\cdots,0}) \nonumber\\
%&\oplus (0, 0, 0, 0, 0, 1, 0, 0; \underline{\pm 1,0,\cdots,0}) \oplus (0, 0, 0, 0, 0, 0, 1, 0; \underline{\pm 1,0,\cdots,0}) \nonumber\\
%&\oplus (0, 0, 0, 0, 0, 0, 0, 1; \underline{\pm 1,0,\cdots,0})\nonumber\\
%120' :~&(0, 0, 0, 0, 0, 0, 0, 0; \underline{\pm 1,\pm 1,\cdots,0})
%\end{align} 

\section{Conclusion}
\label{sec:con}
We have searched for $SO(32)$ heterotic SM vacua directly with the SM gauge group from smooth CY threefolds. 
The non-trivial internal gauge background allows us to directly embed the SM gauge group $G_{\rm SM}$ into $SO(32)$ one. 
In this paper, we focus on the branching $G_{\rm SM}\subset SO(16) \subset SO(32)$ as listed in Table~\ref{tab:group}. 
Against several branchings of $SO(16)$, we have derived a general formula leading to the three generations of quarks and leptons 
and no chiral exotics, taking into account the K-theory condition and the hypercharge masslessness conditions. 
Since the obtained formula is independent of the topological data of CY, it is applicable in general CY compactifications in contrast to models using Wilson lines. 
Such a direct flux breaking is attractive scenario not only in F-theory GUTs~\cite{Beasley:2008kw,Donagi:2008kj}, but also in heterotic string theory.

For concreteness, we have discussed phenomenological aspects of direct flux breaking scenario on a specific CICY, where the spectrum consists of 
MSSM particles, singlets, extra Higgs doublets and vector-like particles whose number cannot be captured by the index theorem. 
In our setup, the normalization of hypercharge direction is different from conventional $SU(5)$ GUT normalization, which also be stated in 
$E_8\times E_8$ heterotic compactification with a hypercharge flux~\cite{Anderson:2014hia}. 
However, one-loop threshold corrections to gauge couplings are non-universal for non-abelian gauge groups compared with $E_8\times E_8$ heterotic models. 
It thus opens up the way for phenomenologically acceptable models. 

It is interesting to check whether or not one can achieve direct flux breaking (i.e., hypercharge flux breaking) in the dual global 
F-theory compactifications.

\section*{Acknowledgments}
  H.~O. would like to thank A. Constantin, T.~Kobayashi and S.~Sugimoto for discussions. 
  H.~O. was supported in part by Grant-in-Aid for Young Scientists (B) (No.~17K14303) 
  from Japan Society for the Promotion of Science.

\appendix 

\section{Group decompositions}
\label{app}

%In this appendix, we show all the group decomposition of $SO(16)\subset SO(32)$ treated in this paper. 
%In particular, we list the order of freely-acting discrete symmetry group $\Gamma$ to realize three-generation 
%of quarks and leptons, taking into account the K-theory condition and hypercharge masslessness conditions. 

In this appendix, we present the algorithm to compute the particle spectrum for 
all the group decompositions of $SO(16)\subset SO(32)$ treated in this paper. 
First we decide which weights correspond to particles or anti-particles 
without determining the hypercharge direction. In our analysis, the matter weights belonging to 
$SO(16)$ adjoint representation are chosen as
\begin{align}
&Q: 
\left\{
\begin{array}{c}
Q_1=(\underline{0,0,-1},\underline{1,0},0,0,0)\\
Q_2=(\underline{0,0,-1},\underline{0,-1},0,0,0)
\end{array}
\right.
,
L: 
\left\{
\begin{array}{c}
L_{1,\pm}=(0, 0, 0, \underline{\pm 1, 0}, \pm 1, 0, 0)\\
L_{2,\pm}=(0, 0, 0, \underline{\pm 1, 0}, \mp 1, 0, 0)\\
L_{3,\pm}=(0, 0, 0, \underline{\pm 1, 0}, 0, \pm 1, 0)\\
L_{4,\pm}=(0, 0, 0, \underline{\pm 1, 0}, 0, \mp 1, 0)\\
L_{5,\pm}=(0, 0, 0, \underline{\pm 1, 0}, 0, 0, \pm 1)\\
L_{6,\pm}=(0, 0, 0, \underline{\pm 1, 0}, 0, 0, \mp 1)\\ 
\end{array}
\right.
,
\nonumber \\
&u^c, d^c: 
\left\{
\begin{array}{c}
u_1^c=(\underline{1, 0, 0}, 0, 0, 1, 0, 0)\\
u_2^c=(\underline{1, 0, 0}, 0, 0, -1, 0, 0)\\
u_3^c=(\underline{1, 0, 0}, 0, 0, 0, 1, 0)\\
u_4^c=(\underline{1, 0, 0}, 0, 0, 0, -1, 0)\\
u_5^c=(\underline{1, 0, 0}, 0, 0, 0, 0,1)\\
u_6^c=(\underline{1, 0, 0}, 0, 0, 0, 0,-1)\\
\end{array}
\right.
,
e^c, {\rm singlets}: 
\left\{
\begin{array}{c}
e_{1,\pm}^c=(0, 0, 0, \pm 1, \pm 1, 0, 0, 0)\\
e_{2,\pm}^c=(0, 0, 0, 0, 0, \pm 1, \pm 1, 0)\\
e_{3,\pm}^c=(0, 0, 0, 0, 0, \mp 1, \pm 1, 0)\\
e_{4,\pm}^c=(0, 0, 0, 0, 0, \pm 1, 0, \pm 1)\\
e_{5,\pm}^c=(0, 0, 0, 0, 0, \mp 1, 0, \pm 1)\\
e_{6,\pm}^c=(0, 0, 0, 0, 0, 0, \pm 1, \pm 1)\\
e_{7,\pm}^c=(0, 0, 0, 0, 0, 0, \mp 1, \pm 1)\\
\end{array}
\right.
,
\nonumber \\
&\mathrm{Exotic\:state:} 
\begin{array}{c}
(\underline{0, -1, -1}, 0, 0, 0, 0, 0)\\
\end{array}
,
\label{eq:weight}
\end{align}
%変更 7/17
(double sign in the same order) under the Cartan direction of $SO(16)$, where the underline represents the 
possible permutations. 
It is now assumed that there are no exotic charged particles except for $(\underline{0, -1, -1}, 0, 0, 0, 0, 0)$. 
Note that when $L_{i,+}$ with positive sign is a left-handed lepton, which has $(1,2)_{-1/2}$ representation in the SM, 
$L_{i,-}$ belonging to $(1,2)_{+1/2}$ representation is an anti left-handed lepton. It holds for $e^c$ and singlets. 
At this stage, we cannot distinguish between $u^c$ and $d^c$ because the hypercharge direction is not specified.

To extract the SM particles from Eq.~(\ref{eq:weight}) including particles and anti-particles, we define the following maps: 
\begin{align}
&q_L:\{1,2,\cdots,6\}\rightarrow\{\pm 1\}, \nonumber\\
&q_{u^c}:\{1,2,\cdots,6\}\rightarrow\{\pm 1\}, \nonumber\\
&q_{e^c}:\{1,2,\cdots,7\}\rightarrow\{\pm 1,\pm 2\}, 
\end{align}
where 
\begin{align}
&q_L(i)=\left\{
\begin{array}{ccl}
+1&&\mathrm{when\:}L_{i,+}\:(L_{i,-})\mathrm{\:is\:the\:left\mathchar`-handed\:(anti\mathchar`-)\:lepton}\\
-1&&\mathrm{when\:}L_{i,-}\:(L_{i,+})\mathrm{\:is\:the\:left\mathchar`-handed\:(anti\mathchar`-)\:lepton}
\end{array}
\right.
,
\nonumber \\
&q_{u^c}(i)=\left\{
\begin{array}{ccl}
+1&&\mathrm{when\:}u_{i}^c\mathrm{\:is\:the\:conjugate\:of\:right\mathchar`-handed\:up\mathchar`-type\:quark}\\
-1&&\mathrm{when\:}u_{i}^c\mathrm{\:is\:the\:conjugate\:of\:right\mathchar`-handed\:down\mathchar`-type\:quark}
\end{array}
\right.
,
\nonumber \\
&q_{e^c}(i)=\left\{
\begin{array}{ccl}
+1&&\mathrm{when\:}e_{i,+}^c\:(e_{i,-}^c)\mathrm{\:is\:the\:conjugate\:of\:right\mathchar`-handed\:(anti\mathchar`-)\:electron}\\
-1&&\mathrm{when\:}e_{i,-}^c\:(e_{i,+}^c)\mathrm{\:is\:the\:conjugate\:of\:right\mathchar`-handed\:(anti\mathchar`-)\:electron}\\
+2&&\mathrm{when\:}e_{i,+}^c\:(e_{i,-}^c)\mathrm{\:is\:the\:conjugate\:of\:right\mathchar`-handed\:(anti\mathchar`-)\:singlet}\\
-2&&\mathrm{when\:}e_{i,-}^c\:(e_{i,+}^c)\mathrm{\:is\:the\:conjugate\:of\:right\mathchar`-handed\:(anti\mathchar`-)\:singlet}\\
\end{array}
\right. .
\end{align}
Now, we have not distinguished between the Higgsino fields and the charged leptons. 

For example, $(q_{u^c}(1),q_{u^c}(2),q_{u^c}(3),q_{u^c}(4),q_{u^c}(5),q_{u^c}(6))=(1,-1,1,-1,-1,1)$ means that 
$u_1^c,u_3^c,u_6^c$ are the conjugates of the right-handed up-type quarks $(\bar{3},1)_{-2/3}$ and $u_2^c,u_4^c,u_5^c$ are 
the conjugates of the right-handed down-type quarks $(\bar{3},1)_{1/3}$.

Next, we searched for all $2^6\cdot 2^6\cdot 4^7\approx 6.71\times 10^7$ possibilities of $\{q_L(i),q_{u^c}(i),q_{e^c}(i)\}$ 
against each branching of $SO(16)$ listed in Table~\ref{tab:group}.
Then we determine the coefficients of hypercharge direction $f_a$ to realize the correct hypercharge of quarks and leptons. 
It is remarkable that, in our analysis, $f_a$ (if exist) are determined uniquely against each set of $\{q_L(i), q_{u^c}(i), q_{e^c}(i), T_a\}$ with 
$U(1)$ generators $T_a$. 
Finally we solve equations for $\{X_{ABC},X'_{AB},X''_A,X''',Z_A,Z'\}$ (like Eqs.~(\ref{eq:indvis}) and (\ref{eq:indhid})) to 
obtain the three generation of quarks and leptons without chiral exotics, taking into account the hypercharge masslessness conditions and the K-theory condition. 
In Table~\ref{tab:group}, we show the possible order of freely-acting discrete symmetry group $\Gamma$. 
In particular, on simply-connected CY threefolds ($|\Gamma|=1$), the list of 24 variables $\{X_{ABC}, X^\prime_{AB}, X^{''}_A, X^{'''}, Z_A, Z^\prime\}$ 
in Table~\ref{tab:hflux} lead to the MSSM-like models, where we choose a specific hypercharge direction for simplicity. 

%of quarks and leptons, taking into account the K-theory condition and hypercharge masslessness conditions. 

In the search of concrete models in Sec.~3, we use the brute force attack to solve Eqs.~(\ref{eq:indvis}) and (\ref{eq:indhid}) within $m_A^{(i)}\in[-1,1]$ for all the possibilities, some of which have solutions of $\{X_{ABC}$, $X'_{AB}$, $X''_A$, $X'''$, $Z_A$, $Z'\}$. 
We check the $D$-term conditions after finding the solution of chiral zero-modes.

\LTcapwidth=\linewidth%captionの幅の調整
\begin{longtable}{|c|c|}
  %\begin{tabular}{|c|c|} 
  \hline
\multicolumn{2}{|c|}{\bf GUT-like decompositions}  \\
\hline \hline
%8
\multicolumn{2}{|l|}{(i)$SO(8)\times U(1)\rightarrow SU(5)\times U(1)^4\rightarrow G_{\rm SM}\times U(1)^4$} 
\\
\hline 
    $U(1)$ generators $T_a$  & Available $|\Gamma|$ 
\\ \hline 
\footnotesize$
     \begin{matrix}
U(1)_1: {\rm diag}(1,1,1,1,1,1,1,1),\quad 
U(1)_2: {\rm diag}(2,2,2,-3,-3,0,0,0),\\
U(1)_3: {\rm diag}(1,1,1,1,1,1,1,-7),\quad
U(1)_4: {\rm diag}(1,1,1,1,1,1,-6,0),\\
U(1)_5: {\rm diag}(1,1,1,1,1,-5,0,0),\\
\end{matrix}
\footnotesize$          & $|\Gamma|=10\mathbb{Z}_{>0}$
      \\ \hline      
%71
      \multicolumn{2}{|l|}{(ii)$SO(14)\times SO(2)\rightarrow SU(7)\times U(1)^2\rightarrow SU(5)\times U(1)^4\rightarrow G_{\rm SM}\times U(1)^4$}  \\
\hline
    $U(1)$ generators $T_a$ & Available $|\Gamma|$ 
\\ \hline 
\footnotesize$
     \begin{matrix}
U(1)_1: {\rm diag}(1,1,1,1,1,1,1,0),\quad
U(1)_2: {\rm diag}(1,1,1,1,1,1,-6,0)\\
U(1)_3: {\rm diag}(1,1,1,1,1,-5,0,0),
U(1)_4: {\rm diag}(0,0,0,0,0,0,0,1)\\
U(1)_5: {\rm diag}(2,2,2,-3,-3,0,0,0)\\
\end{matrix}
\footnotesize$          & $|\Gamma|=5\mathbb{Z}_{>0}$
      \\ \hline
%611
      \multicolumn{2}{|l|}{(iii)$SO(12)\times SO(2)^2\rightarrow SU(6)\times U(1)^3\rightarrow SU(5)\times U(1)^4\rightarrow G_{\rm SM}\times U(1)^4$}  \\
\hline
    $U(1)$ generators $T_a$ & Available $|\Gamma|$ 
\\ \hline 
\hspace{-30pt}(a)
\footnotesize$
     \begin{matrix}
U(1)_1: {\rm diag}(0, 0, 0, 0, 0, 0, 1, 0),\quad
U(1)_2: {\rm diag}(2, 2, 2, -3, -3, 0, 0, 0)\\
U(1)_3: {\rm diag}(1, 1, 1, 1, 1, 1, 0, 0),\quad
U(1)_4: {\rm diag}(1, 1, 1, 1, 1, -5, 0, 0)\\
U(1)_5: {\rm diag}(0, 0, 0, 0, 0, 0, 0, 1)\\
\end{matrix}
\footnotesize$         & $|\Gamma|=5\mathbb{Z}_{>0}$
      \\ \hline  
%62
\hspace{-30pt}(b)
\footnotesize$
     \begin{matrix}
U(1)_1: {\rm diag}(0, 0, 0, 0, 0, 0, 1, 1),\quad
U(1)_2: {\rm diag}(2, 2, 2, -3, -3, 0, 0, 0)\\
U(1)_3: {\rm diag}(1, 1, 1, 1, 1, 1, 0, 0),\quad
U(1)_4: {\rm diag}(1, 1, 1, 1, 1, -5, 0, 0)\\
U(1)_5: {\rm diag}(0, 0, 0, 0, 0, 0, 1, -1)\\
\end{matrix}
\footnotesize$ & $|\Gamma|=5\mathbb{Z}_{>0}$
      \\ \hline  
%5111
      \multicolumn{2}{|l|}{(iv)$SO(10)\times SO(2)^3\rightarrow SU(5)\times U(1)^4\rightarrow G_{\rm SM}\times U(1)^4$}  \\
\hline
    $U(1)$ generators $T_a$  & Available $|\Gamma|$ 
\\ \hline 
\footnotesize$
     \begin{matrix}
U(1)_1: {\rm diag}(1,1,1,1,1,1,1,1),\quad
U(1)_2: {\rm diag}(0,0,0,0,0,0,0,1)\\
U(1)_3: {\rm diag}(2,2,2,-3,-3,0,0,0),\quad
U(1)_4: {\rm diag}(0,0,0,0,0,0,1,0)\\
U(1)_5: {\rm diag}(0,0,0,0,0,1,0,0)\\
\end{matrix}
\footnotesize$       & $|\Gamma|=5\mathbb{Z}_{>0}$
      \\ \hline
%521          
           \multicolumn{2}{|l|}{(v)$SO(10)\times SO(2)\times SO(4)\rightarrow SU(5)\times U(1)^4\rightarrow G_{\rm SM}\times U(1)^4$}  \\
\hline
    $U(1)$ generators $T_a$  & Available $|\Gamma|$ 
\\ \hline 
\footnotesize$
     \begin{matrix}
U(1)_1: {\rm diag}(1, 1, 1, 1, 1, 0, 0, 0),\quad
U(1)_2: {\rm diag}(0,0,0,0,0,0,0,1)\\
U(1)_3: {\rm diag}(2,2,2,-3,-3,0,0,0),\quad
U(1)_4: {\rm diag}(0, 0, 0, 0, 0, 1, 1, 0)\\
U(1)_5: {\rm diag}(0, 0, 0, 0, 0, 1, -1, 0)\\
\end{matrix}
\footnotesize$         & $|\Gamma|=5\mathbb{Z}_{>0}$
      \\ \hline
%53          
                 \multicolumn{2}{|l|}{(vi)$SO(10)\times SO(6)\rightarrow SU(5)\times SU(3)\times U(1)^2\rightarrow G_{\rm SM}\times U(1)^4$}  \\
\hline
    $U(1)$ generators $T_a$  & Available $|\Gamma|$ 
\\ \hline 
\hspace{-30pt}(a)
\footnotesize$
     \begin{matrix}
U(1)_1: {\rm diag}(1, 1, 1, 1, 1, 0, 0, 0),\quad
U(1)_2: {\rm diag}(2, 2, 2, -3, -3, 0, 0, 0)\\
U(1)_3: {\rm diag}(0, 0, 0, 0, 0, 1, 1, 1),\quad
U(1)_4: {\rm diag}(0, 0, 0, 0, 0, 1, 1, -2)\\
U(1)_5: {\rm diag}(0, 0, 0, 0, 0, 1, -1, 0)\\
\end{matrix}
\footnotesize$        & $|\Gamma|=\mathbb{Z}_{>0}$
%53.2      
\\ \hline 
\hspace{-30pt}(b)
\footnotesize$
     \begin{matrix}
U(1)_1: {\rm diag}(1, 1, 1, 0, 0, 0, 1, 1),\quad
U(1)_2: {\rm diag}(0, 0, 0, 1, 1, 1, 0, 0)\\
U(1)_3: {\rm diag}(0, 0, 0, 1, 1, -2, 0, 0),\quad
U(1)_4: {\rm diag}(1, 1, 1, 0, 0, 0, 1, -4)\\
U(1)_5: {\rm diag}(1, 1, 1, 0, 0, 0, -3, 0)\\
\end{matrix}
\footnotesize$        & $|\Gamma|=\mathbb{Z}_{>0}$
      \\ \hline
%7521         
            \multicolumn{2}{|l|}{(vii)$SO(14)\times SO(2)\rightarrow SU(5)\times SU(2)_L\times U(1)^3\rightarrow G_{\rm SM}\times U(1)^4$}  \\
\hline
    $U(1)$ generators $T_a$ & Available $|\Gamma|$ 
\\ \hline 
\footnotesize$
     \begin{matrix}
U(1)_1: {\rm diag}(1,1,1,1,1,1,1,0),\quad
U(1)_2: {\rm diag}(2,2,2,-3,-3,0,0,0)\\
U(1)_3: {\rm diag}(1, 1, 1, 1, 1, -\frac{5}{2}, -\frac{5}{2}, 0),\quad
U(1)_4: {\rm diag}(0,0,0,0,0,0,0,1)\\
U(1)_5: {\rm diag}(0, 0, 0, 0, 0, 1, -1, 0)\\
\end{matrix}
\footnotesize$         & $|\Gamma|=5\mathbb{Z}_{>0}$
      \\ \hline  
%87521
                  \multicolumn{2}{|l|}{(viii)$SU(8)\times U(1)\rightarrow SU(7)\times U(1)^2\rightarrow SU(5)\times SU(2)_L\times U(1)^3\rightarrow G_{\rm SM}\times U(1)^4$}  \\
\hline
    $U(1)$ generators $T_a$  & Available $|\Gamma|$ 
\\ \hline 
\hspace{-21pt}(a)
\footnotesize$
     \begin{matrix}
U(1)_1: {\rm diag}(1,1,1,1,1,1,1,1),\quad
U(1)_2: {\rm diag}(2,2,2,-3,-3,0,0,0)\\
U(1)_3: {\rm diag}(1, 1, 1, 1, 1, -\frac{5}{2}, -\frac{5}{2}, 0),\quad
U(1)_4: {\rm diag}(1,1,1,1,1,1,1,-7)\\
U(1)_5: {\rm diag}(0, 0, 0, 0, 0, 1, -1, 0)\\
\end{matrix}
\footnotesize$         & $|\Gamma|=5\mathbb{Z}_{>0}$
      \\ \hline   
%862
\hspace{-23pt}(b)
\footnotesize$
     \begin{matrix}
U(1)_1: {\rm diag}(1,1,1,1,1,1,1,1),\quad
U(1)_2: {\rm diag}(2,2,2,-3,-3,0,0,0)\\
U(1)_3: {\rm diag}(1, 1, 1, 1, 1, 1, -3, -3),\quad
U(1)_4: {\rm diag}(1, 1, 1, 1, 1, -5, 0, 0)\\
U(1)_5: {\rm diag}(0, 0, 0, 0, 0, 0, -1, 1)\\
\end{matrix}
\footnotesize$         & $|\Gamma|=5\mathbb{Z}_{>0}$
      \\ \hline
%853
\hspace{-5pt}(c)
\footnotesize$
     \begin{matrix}
U(1)_1: {\rm diag}(1,1,1,1,1,1,1,1),\quad
U(1)_2: {\rm diag}(2,2,2,-3,-3,0,0,0)\\
U(1)_3: {\rm diag}(\frac{3}{2}, \frac{3}{2}, \frac{3}{2}, \frac{3}{2}, \frac{3}{2}, -\frac{5}{2}, -\frac{5}{2}, -\frac{5}{2}),\quad
U(1)_4: {\rm diag}(0, 0, 0, 0, 0, 1, 1, -2)\\
U(1)_5: {\rm diag}(0, 0, 0, 0, 0, 1, -1, 0)\\
\end{matrix}
\footnotesize$        & $|\Gamma|=5\mathbb{Z}_{>0}$
\\ \hline
%8.62
                  \multicolumn{2}{|l|}{(ix)$SU(8)\times U(1)\rightarrow SU(6)\times SU(2)\times U(1)^2\rightarrow SU(5)\times SU(2)_L\times U(1)^3\rightarrow G_{\rm SM}\times U(1)^4$}  \\
\hline
    $U(1)$ generators $T_a$  & Available $|\Gamma|$ 
\\ \hline 
\footnotesize$
     \begin{matrix}
U(1)_1: {\rm diag}(1, 1, 1, 1, 1, 1, 1, 1),\quad
U(1)_2: {\rm diag}1, 1, 1, 1, 1, 1, -3, -3)\\
U(1)_3: {\rm diag}(1, 1, 1, 1, 1, -5, 0, 0),\quad
U(1)_4: {\rm diag}(0, 0, 0, 0, 0, 0, 1, -1)\\
U(1)_5: {\rm diag}(2, 2, 2, -3, -3, 0, 0, 0)\\
\end{matrix}
\footnotesize$         & $|\Gamma|=5\mathbb{Z}_{>0}$
      \\ \hline     \hline  
      \multicolumn{2}{|c|}{\bf Pati-Salam-like decompositions}  \\
\hline \hline
%422
\multicolumn{2}{|l|}{(x)$SO(8)\times SO(4)\times SO(4)\rightarrow SU(4)_C\times SU(2)_L\times SU(2)\times U(1)^3\rightarrow G_{\rm SM}\times U(1)^4$}  
\\
\hline 
    $U(1)$ generators $T_a$ & Available $|\Gamma|$ 
\\ \hline 
\footnotesize$
     \begin{matrix}
U(1)_1: {\rm diag}(1, 1, 1, 0, 0, 0, 0, 1),\quad
U(1)_2: {\rm diag}(1, 1, 1, 0, 0, 0, 0, -3)\\
U(1)_3: {\rm diag}(0, 0, 0, 1, 1, 0, 0, 0),\quad
U(1)_4: {\rm diag}(0, 0, 0, 0, 0, 1, 1, 0)\\
U(1)_5: {\rm diag}(0, 0, 0, 0, 0, 1, -1, 0)\\
\end{matrix}
\footnotesize$          & $|\Gamma|=\mathbb{Z}_{>0}$
      \\ \hline    
%44
      \multicolumn{2}{|l|}{(xi)$SU(4)\times SU(4)\times U(1)^2\rightarrow SU(4)_C\times SU(2)_L\times SU(2)\times U(1)^3\rightarrow G_{\rm SM}\times U(1)^4$}  
\\
\hline 
    $U(1)$ generators $T_a$ & Available $|\Gamma|$ 
\\ \hline 
\footnotesize$
     \begin{matrix}
U(1)_1: {\rm diag}(1, 1, 1, 0, 0, 0, 0, 1),\quad
U(1)_2: {\rm diag}(1, 1, 1, 0, 0, 0, 0, -3)\\
U(1)_3: {\rm diag}(0, 0, 0, 1, 1, -1, -1, 0),\quad
U(1)_4: {\rm diag}(0, 0, 0, 1, 1, 1, 1, 0)\\
U(1)_5: {\rm diag}(0, 0, 0, 0, 0, 1, -1, 0)\\
\end{matrix}
\footnotesize$        & $|\Gamma|=\mathbb{Z}_{>0}$
      \\ \hline     \hline  
      %%%%%%%%%%%%%%%%%%%%%%%%%%%%%%%%%%%%%%%%%%%%%%%%%%%%%%%%%%%%%%%%%%%%%%%%%%%%%%%%%%%%%%%%%%%%%%%%%%%%%%%%%%%%%%%
      \multicolumn{2}{|c|}{\bf Others}  \\
\hline \hline
\multicolumn{2}{|l|}{(xii)$SO(6)\times SO(6)\times SO(4)\rightarrow SU(3)_C\times SU(3)\times SU(2)\times U(1)^3\rightarrow G_{\rm SM}\times U(1)^4$}  
%332
\\
\hline 
    $U(1)$ generators $T_a$ & Available $|\Gamma|$ 
\\ \hline
\hspace{-38pt}(a) 
\footnotesize$
     \begin{matrix}
U(1)_1: {\rm diag}(0, 0, 0, 0, 0, 0, 1, 1),\quad
U(1)_2: {\rm diag}(0, 0, 0, 1, 1, 1, 0, 0)\\
U(1)_3: {\rm diag}(0, 0, 0, 1, 1, -2, 0, 0),\quad
U(1)_4: {\rm diag}(1, 1, 1, 0, 0, 0, 0, 0)\\
U(1)_5: {\rm diag}(0, 0, 0, 0, 0, 0, 1, -1)\\
\end{matrix}
\footnotesize$         & $|\Gamma|=\mathbb{Z}_{>0}$
%323
\\ \hline 
\hspace{-30pt}(b)
\footnotesize$
     \begin{matrix}
U(1)_1: {\rm diag}(1, 1, 1, 0, 0, 0, 0, 0),\quad
U(1)_2: {\rm diag}(0, 0, 0, 1, 1, 0, 0, 0)\\
U(1)_3: {\rm diag}(0, 0, 0, 0, 0, 1, 1, -2),\quad
U(1)_4: {\rm diag}(0, 0, 0, 0, 0, 1, -1, 0)\\
U(1)_5: {\rm diag}(0, 0, 0, 0, 0, 1, 1, 1)\\
\end{matrix}
\footnotesize$         & $|\Gamma|=\mathbb{Z}_{>0}$
      \\ \hline    
%32111
      \multicolumn{2}{|l|}{(xiii)$SO(6)\times SO(4)\times SO(2)^3\rightarrow G_{\rm SM}\times U(1)^4$}  
\\
\hline 
    $U(1)$ generators $T_a$  & Available $|\Gamma|$ 
\\ \hline 
\footnotesize$
     \begin{matrix}
U(1)_1: {\rm diag}(1, 1, 1, 0, 0, 0, 0, 1),\quad
U(1)_2: {\rm diag}(0, 0, 0, 1, 1, 0, 0, 0)\\
U(1)_3: {\rm diag}(0, 0, 0, 0, 0, 1, 0, 0),\quad
U(1)_4: {\rm diag}(0, 0, 0, 0, 0, 0, 1, 0)\\
U(1)_5: {\rm diag}(0, 0, 0, 0, 0, 0, 0, 1)\\
\end{matrix}
\footnotesize$          &  $|\Gamma|=\mathbb{Z}_{>0}$
      \\ \hline   
%431      
            \multicolumn{2}{|l|}{(xiv)$SO(8)\times SO(6)\times SO(2)\rightarrow SU(4)_C\times SU(3)\times U(1)^3\rightarrow G_{\rm SM}\times U(1)^4$}  
\\
\hline 
    $U(1)$ generators $T_a$  & Available $|\Gamma|$ 
\\ \hline 
\hspace{-38pt}(a)
\footnotesize$
     \begin{matrix}
U(1)_1: {\rm diag}(1, 1, 1, 0, 0, 0, 0, 1),\quad
U(1)_2: {\rm diag}(0, 0, 0, 1, 1, 1, 0, 0)\\
U(1)_3: {\rm diag}(0, 0, 0, 1, 1, -2, 0, 0),\quad
U(1)_4: {\rm diag}(0, 0, 0, 0, 0, 0, 1, 0)\\
U(1)_5: {\rm diag}(1, 1, 1, 0, 0, 0, 0, -3)\\
\end{matrix}
\footnotesize$          &  $|\Gamma|=\mathbb{Z}_{>0}$
%341           
\\ \hline 
\hspace{-30pt}(b)
\footnotesize$
     \begin{matrix}
U(1)_1: {\rm diag}(1, 1, 1, 0, 0, 0, 0, 0),\quad
U(1)_2: {\rm diag}(0, 0, 0, 1, 1, 1, 1, 0)\\
U(1)_3: {\rm diag}(0, 0, 0, 1, 1, 1, -3, 0),\quad
U(1)_4: {\rm diag}(0, 0, 0, 1, 1, -2, 0, 0)\\
U(1)_5: {\rm diag}(0, 0, 0, 0, 0, 0, 0, 1)\\
\end{matrix}
\footnotesize$          &  $|\Gamma|=\mathbb{Z}_{>0}$
      \\ \hline
%341.2           
\hspace{-23pt}(c)
\footnotesize$
     \begin{matrix}
U(1)_1: {\rm diag}(1, 1, 1, 0, 0, 0, 0, 0),\quad
U(1)_2: {\rm diag}(0, 0, 0, 1, 1, 1, 1, 0)\\
U(1)_3: {\rm diag}(0, 0, 0, 1, 1, -1, -1, 0),\quad
U(1)_4: {\rm diag}(0, 0, 0, 0, 0, 1, -1, 0)\\
U(1)_5: {\rm diag}(0, 0, 0, 0, 0, 0, 0, 1)\\
\end{matrix}
\footnotesize$          &  $|\Gamma|=\mathbb{Z}_{>0}$
      \\ \hline
%4211              
            \multicolumn{2}{|l|}{(xv)$SO(8)\times SO(4)\times SO(2)^2\rightarrow SU(4)_C\times SU(2)_L\times U(1)^3\rightarrow G_{\rm SM}\times U(1)^4$}  
\\
\hline 
    $U(1)$ generators $T_a$  & Available $|\Gamma|$ 
\\ \hline 
\footnotesize$
     \begin{matrix}
U(1)_1: {\rm diag}(1, 1, 1, 0, 0, 0, 0, 1),\quad
U(1)_2: {\rm diag}(0, 0, 0, 0, 0, 0, 1, 0)\\
U(1)_3: {\rm diag}(0, 0, 0, 1, 1, 0, 0, 0),\quad
U(1)_4: {\rm diag}(0, 0, 0, 0, 0, 1, 0, 0)\\
U(1)_5: {\rm diag}(1, 1, 1, 0, 0, 0, 0, -3)\\
\end{matrix}
\footnotesize$          &  $|\Gamma|=\mathbb{Z}_{>0}$
      \\ \hline   
%35           
            \multicolumn{2}{|l|}{(xvi)$SO(6)\times SO(10)\rightarrow SU(3)_C\times SU(5)\times U(1)^2\rightarrow G_{\rm SM}\times U(1)^4$}  
\\
\hline 
    $U(1)$ generators $T_a$  & Available $|\Gamma|$ 
\\ \hline
\hspace{-31pt}(a) 
\footnotesize$
     \begin{matrix}
U(1)_1: {\rm diag}(1, 1, 1, 0, 0, 0, 0, 0),\quad
U(1)_2: {\rm diag}(0, 0, 0, 1, 1, 1, 1, 1)\\
U(1)_3: {\rm diag}(0, 0, 0, 1, 1, 1, 1, -4),\quad
U(1)_4: {\rm diag}(0, 0, 0, 1, 1, 1, -3, 0)\\
U(1)_5: {\rm diag}(0, 0, 0, 1, 1, -2, 0, 0)\\
\end{matrix}
\footnotesize$          &  $|\Gamma|=\mathbb{Z}_{>0}$
      \\ \hline
%35.2     
\hspace{-15pt}(b)
\footnotesize$
     \begin{matrix}
U(1)_1: {\rm diag}(1, 1, 1, 0, 0, 0, 0, 0),\quad
U(1)_2: {\rm diag}(0, 0, 0, 1, 1, 1, 1, 1)\\
U(1)_3: {\rm diag}(0, 0, 0, 3, 3, -2, -2, -2),\quad
U(1)_4: {\rm diag}(0, 0, 0, 0, 0, 1, 1, -2)\\
U(1)_5: {\rm diag}(0, 0, 0, 0, 0, 1, -1, 0)\\
\end{matrix}
\footnotesize$          &  $|\Gamma|=\mathbb{Z}_{>0}$
      \\ \hline
%35.3     
\hspace{-23pt}(c)
\footnotesize$
     \begin{matrix}
U(1)_1: {\rm diag}(1, 1, 1, 0, 0, 0, 0, 0),\quad
U(1)_2: {\rm diag}(0, 0, 0, 1, 1, 1, 1, 1)\\
U(1)_3: {\rm diag}(0, 0, 0, 2, 2, 2, -3, -3),\quad
U(1)_4: {\rm diag}(0, 0, 0, 1, 1, -2, 0, 0)\\
U(1)_5: {\rm diag}(0, 0, 0, 0, 0, 0, 1, -1)\\
\end{matrix}
\footnotesize$          &  $|\Gamma|=\mathbb{Z}_{>0}$
      \\ \hline
%3221           
            \multicolumn{2}{|l|}{(xvii)$SO(6)\times SO(4)\times SO(4)\times SO(2)\rightarrow SU(3)_C\times SU(2)_L\times SU(2)\times U(1)^4\rightarrow G_{\rm SM}\times U(1)^4$}  
\\
\hline 
    $U(1)$ generators $T_a$  & Available $|\Gamma|$ 
\\ \hline 
\footnotesize$
     \begin{matrix}
U(1)_1: {\rm diag}(1, 1, 1, 0, 0, 0, 0, 0),\quad
U(1)_2: {\rm diag}(0, 0, 0, 1, 1, 0, 0, 0)\\
U(1)_3: {\rm diag}(0, 0, 0, 0, 0, 1, 1, 0),\quad
U(1)_4: {\rm diag}(0, 0, 0, 0, 0, 0, 0, 1)\\
U(1)_5: {\rm diag}(0, 0, 0, 0, 0, 1, -1, 0)\\
\end{matrix}
\footnotesize$          &  $|\Gamma|=\mathbb{Z}_{>0}$
      \\ \hline
%3311           
            \multicolumn{2}{|l|}{(xviii)$SO(6)\times SO(6)\times SO(2)^2\rightarrow SU(3)_C\times SU(3)\times U(1)^4\rightarrow G_{\rm SM}\times U(1)^4$}  
\\
\hline 
    $U(1)$ generators $T_a$  & Available $|\Gamma|$ 
\\ \hline 
\footnotesize$
     \begin{matrix}
U(1)_1: {\rm diag}(1, 1, 1, 0, 0, 0, 0, 0),\quad
U(1)_2: {\rm diag}(0, 0, 0, 0, 0, 0, 1, 0)\\
U(1)_3: {\rm diag}(0, 0, 0, 1, 1, -2, 0, 0),\quad
U(1)_4: {\rm diag}(0, 0, 0, 1, 1, 1, 0, 0)\\
U(1)_5: {\rm diag}(0, 0, 0, 0, 0, 0, 0, 1)\\
\end{matrix}
\footnotesize$          &  $|\Gamma|=\mathbb{Z}_{>0}$
      \\ \hline                                           
 % \end{tabular}
  \caption{Group decomposition of $SO(16)$ discussed in this paper. In all cases, the hypercharge direction is 
  $U(1)_Y=\frac{1}{6}{\rm diag}(-1,-1,-1,0,0,\pm 3, \pm 3, \pm 3)$ where the sign is taken in the random order. $\Gamma$ denotes 
  the possible order of freely-acting discrete symmetry group of CY threefold, satisfying the phenomenological requirements~(\ref{eq:indvis}) 
  and (\ref{eq:indhid}).}
  \label{tab:group}
\end{longtable}

%%%%%%%%%%%%%%%%%%%%%%%%%%%%%%%%%%%%%%%%%%%%%%%%%%%%%%%%%%%%%%%%%%%%%%%%%%%%%%%%%%%%%%%%%%%%%%%%%%%%%%%%%%%%%%%%%%

\begin{longtable}{|l|}
  %\begin{tabular}{|c|c|} 
  %\hline
  \hline
        \multicolumn{1}{|c|}{\bf GUT-like decompositions}  \\ 
\hline 
%53
\multicolumn{1}{|l|}{(vi)-(a)$SO(10)\times SO(6)\rightarrow SU(5)\times SU(3)\times U(1)^2\rightarrow G_{\rm SM}\times U(1)^4$}  
\\
\multicolumn{1}{|l|}{\underline{Hypercharge direction: $U(1)_Y=\left(-3U(1)_1-U(1)_2+15U(1)_3\right)/30$}}  
\\
\footnotesize
$X_{333}=0, X_{334}=p_1, X_{335}=p_2, X_{344}=p_3, X_{345}=p_4, X_{355}=-1-3p_3, X_{444}=p_5$,
$X_{445}=p_6$, 
\\
\footnotesize
$X_{455}=1+p_5, X_{555}=p_7, X^\prime_{33}=p_8, X^\prime_{34}=p_{10}, X^\prime_{35}=p_{11}$,  
$X^\prime_{44}=p_{12},  X^\prime_{45}=p_{13}, X^\prime_{55}=p_{14}, X^{''}_{3}=1+12p_{8}$,
\\
\footnotesize
$X^{''}_{4}=p_{15}$, $X^{''}_{5}=p_{16}$, 
$X^{'''}=p_{9}, Z_3=0,  Z_4=-6p_1+12p_3-8p_5, Z_5=-6p_2-12p_4-6p_6-2p_7$, 
\\
\footnotesize
$Z^{'}=36+216 p_8-2p_9$,
      \\ \hline
%53.2  
\multicolumn{1}{|l|}{(vi)-(b)$SO(10)\times SO(6)\rightarrow SU(5)\times SU(3)\times U(1)^2\rightarrow G_{\rm SM}\times U(1)^4$}  
\\
\multicolumn{1}{|l|}{\underline{Hypercharge direction: $U(1)_Y=\left(-3U(1)_1+5U(1)_2-5U(1)_3+3U(1)_4-5U(1)_5\right)/30$}}  
\\
\footnotesize
$X_{333}=p_1, X_{334}=p_2, X_{335}=p_3, X_{344}=p_4, X_{345}=p_5, X_{355}=p_6, X_{444}=p_7$,
$X_{445}=p_8, X_{455}=p_9$, 
\\
\footnotesize
$X_{555}=5p_{10}, X^\prime_{33}=p_1+2p_{11}, X^\prime_{34}=p_{12}, X^\prime_{35}=p_{1}+p_{11}+p_{12}+2p_{13}$,  
$X^\prime_{44}=p_{14}$, 
\\
\footnotesize
$X^\prime_{45}=3p_4+3p_5+6 p_7+6 p_8+10 p_9+6 p_{10}+3 p_{14}+12 p_{15}$, 
\\
\footnotesize
$X^\prime_{55}=5 p_4+3 p_6+10 p_7+10 p_8+10 p_9+30 p_{10}+5 p_{14}+20 p_{15}$, 
\\
\footnotesize
$X^{''}_{3}=144+p_1+100 p_2+36 p_3+100 p_4+40 p_5+20 p_6+80 p_8+80 p_{10}+152 p_{11}+152 p_{12}+144 p_{13}+160 p_{16}$,
\\
\footnotesize
$X^{''}_{4}=24+15 p_2+48 p_4+24 p_5+36 p_7+8 p_8+84 p_9+96 p_{10}+24 p_{11}+30 p_{12}+24 p_{13}+8 p_{14}+72 p_{15}+24 p_{16}$, 
\\
\footnotesize
$X^{''}_{5}=-96+6 p_1-120 p_2-9 p_3-120 p_5-60 p_8+20 p_{10}-114 p_{11}-114 p_{12}-108 p_{13}-120 p_{16}$, 
\\
\footnotesize
$X^{'''}=9 p_1-540 p_2+324 p_3+660 p_4-1080 p_5+180 p_6-1480 p_7+1320 p_8-1560 p_9-1640 p_{10}-126 p_{11}$
\\
\footnotesize
\hspace{25pt}$-72 p_{12}-144 p_{13}-240 p_{14}-960 p_{15}$, 
\\
\footnotesize
$Z_3=-54-18 p_1-15 p_2-36 p_3-75 p_4+60 p_5-45 p_6-30 p_8-30 p_{10}-69 p_{11}-57 p_{12}-69 p_{13}-60 p_{16}$,
\\
\footnotesize
$Z_4=-9-9 p_2-9 p_4-9 p_5-8 p_7+15 p_8-18 p_9+3 p_{10}-9 p_{11}-9 p_{12}-9 p_{13}+6 p_{14}+9 p_{15}-9 p_{16}$, 
\\
\footnotesize
$Z_5=36+45 p_2+15 p_4+45 p_5+30 p_7+15 p_8+75 p_9-55 p_{10}+45 p_{11}+45 p_{12}+45 p_{13}+15 p_{14}+60 p_{15}+45 p_{16}$, 
 \\
\footnotesize
$Z^{'}=-54 p_1+135 p_2-108 p_3-105 p_4+360 p_5-135 p_6+140 p_7-210 p_8+330 p_9-230 p_{10}-27 p_{11}+9 p_{12}$
\\
\footnotesize
\hspace{25pt}$-27 p_{13}-30 p_{14}+30 p_{15}$,
      \\ \hline \hline  
%%%%%%%%%%%%%%%%%%%%%%%%%%%%%%%%%%%%%%%%%%%%%%%%%%%%%%%%%%%%%%%%%%%%%%%%%%%%%%%%%%%%%%%%%%%%%%%%%%%%%%%%%%%%%
      \multicolumn{1}{|c|}{\bf Pati-Salam-like decompositions}  \\ \hline
%422
\multicolumn{1}{|l|}{(x)$SO(8)\times SO(4)\times SO(4)\rightarrow SU(4)_C\times SU(2)_L\times SU(2)\times U(1)^3\rightarrow G_{\rm SM}\times U(1)^4$}  
\\ 
\multicolumn{1}{|l|}{\underline{Hypercharge direction: $U(1)_Y=\left(-U(1)_2+3U(1)_4\right)/6$}}  
\\
\footnotesize
$X_{333}=p_1, X_{334}=p_2, X_{335}=p_{12}, X_{344}=p_3, X_{345}=p_{13}, X_{355}=p_4, X_{444}=p_5$,
$X_{445}=p_6, X_{455}=p_7$, 
\\
\footnotesize
$X_{555}=p_{8}, X^\prime_{33}=6+p_1, X^\prime_{34}=p_{9}, X^\prime_{35}=p_{14}$, 
$X^\prime_{44}=p_{10},  X^\prime_{45}=p_{15}$, 
\\
\footnotesize
$X^\prime_{55}=-6-p_3+p_4-4 p_5+4 p_7+p_{10}$, 
$X^{''}_{3}=p_{11}$,
$X^{''}_{4}=-p_2-4 p_3-4 p_5+4 p_7+2 p_9+4 p_{10}$, 
\\
\footnotesize
$X^{''}_{5}=p_{16}$, 
$X^{'''}=-18-2 p_1+3 p_{11}$, 
$Z_3=-2 p_1$,
$Z_4=-2 p_5-6 p_7$, 
$Z_5=-6 p_6-2 p_8$, 
$Z^{'}=-2 p_1$,
      \\ \hline
%44  
\multicolumn{1}{|l|}{(xi)$SU(4)\times SU(4)\times U(1)^2\rightarrow SU(4)\times SU(2)\times SU(2)\times U(1)^3\rightarrow G_{\rm SM}\times U(1)^4$}  
\\
\multicolumn{1}{|l|}{\underline{Hypercharge direction: $U(1)_Y=\left(-2U(1)_2-3U(1)_3+3U(1)_4\right)/12$}}  
\\
\footnotesize
$X_{333}=p_1, X_{334}=p_2, X_{335}=p_{3}, X_{344}=p_4, X_{345}=p_5, X_{355}=p_6, X_{444}=p_7$,
$X_{445}=p_8, X_{455}=p_9$, 
\\
\footnotesize
$X_{555}=p_{10}, X^\prime_{33}=p_{11}, X^\prime_{34}=p_{12}, X^\prime_{35}=p_{14}$, 
$X^\prime_{44}=6+p_1+3 p_2+3 p_4+p_7-p_{11}-2 p_{12},  X^\prime_{45}=p_{15}$, 
\\
\footnotesize
$X^\prime_{55}=4 p_1-8 p_2+16 p_4-3 p_6-4 p_7+5 p_9-4 p_{12}$, 
$X^{''}_{3}=p_{13}$,
\\
\footnotesize
$X^{''}_{4}=36+7 p_1+11 p_2+33 p_4-4 p_6-3 p_7+4 p_9-4 p_{11}-20 p_{12}+p_{13}$, 
$X^{''}_{5}=p_{16}$, 
\\
\footnotesize
$X^{'''}=90+19 p_1+27 p_2+93 p_4-12 p_6-11 p_7+12 p_9-12 p_{11}-60 p_{12}+6 p_{13}$, 
\\
\footnotesize
$Z_3=-2 p_1-6 p_4-3 p_6+3 p_9$,
$Z_4=-6 p_2+3 p_6-2 p_7-3 p_9$, 
\\
\footnotesize
$Z_5=-6 p_3+12 p_5-6 p_8-2 p_{10}$, 
$Z^{'}=-2 p_1-6 p_2-6 p_4-2 p_7$,
      \\ \hline \hline
%%%%%%%%%%%%%%%%%%%%%%%%%%%%%%%%%%%%%%%%%%%%%%%%%%%%%%%%%%%%%%%%%%%%%%%%%%%%%%%%%%%%%%%%%%%%%%%%%%%%%%%%%%%%%
      \multicolumn{1}{|c|}{\bf Others}  \\ \hline
%332
\multicolumn{1}{|l|}{(xii)-(a)$SO(6)\times SO(6)\times SO(4)\rightarrow SU(3)_C\times SU(3)\times SU(2)\times U(1)^3\rightarrow G_{\rm SM}\times U(1)^4$}  
\\
\multicolumn{1}{|l|}{\underline{Hypercharge direction: $U(1)_Y=\left(3U(1)_1+U(1)_2-U(1)_3-U(1)_4\right)/6$}}  
\\
\footnotesize
$X_{333}=p_1, X_{334}=1+p_2, X_{335}=p_{3}, X_{344}=p_4, X_{345}=p_5, X_{355}=p_6, X_{444}=0$,
$X_{445}=p_7, X_{455}=p_8$, 
\\
\footnotesize
$X_{555}=p_9, X^\prime_{33}=p_{10}, X^\prime_{34}=p_{11}, X^\prime_{35}=p_{12}$, 
$X^\prime_{44}=24+27 p_2+21 p_4+2 p_8-9 p_{11},  X^\prime_{45}=p_{13}$, 
$X^\prime_{55}=p_{14}$, 
\\
\footnotesize
$X^{''}_{3}=p_{15}$,
$X^{''}_{4}=99+108 p_2+84 p_4+8 p_8-36 p_{11}$, 
$X^{''}_{5}=p_3+p_7+2 p_{16}$, 
\\
\footnotesize
$X^{'''}=372+27 p_1+441 p_2+294 p_4-12 p_6+14 p_8-27 p_{10}-147 p_{11}+4 p_{14}+9 p_{15}$, 
\\
\footnotesize
$Z_3=-36-18 p_1-72 p_2-24 p_4+18 p_{10}+24 p_{11}-6 p_{15}$,
$Z_4=0$, 
\\
\footnotesize
$Z_5=-15 p_3-27 p_5-15 p_7-2 p_9+9 p_{12}+9 p_{13}-3 p_{16}$, 
\\
\footnotesize
$Z^{'}=-132-54 p_1-234 p_2-84 p_4+24 p_6+20 p_8+54 p_{10}+78 p_{11}-8 p_{14}-18 p_{15}$,
      \\ \hline
%323
      \multicolumn{1}{|l|}{(xii)-(b)$SO(6)\times SO(6)\times SO(4)\rightarrow SU(3)_C\times SU(3)\times SU(2)\times U(1)^3\rightarrow G_{\rm SM}\times U(1)^4$}  
\\
\multicolumn{1}{|l|}{\underline{Hypercharge direction: $U(1)_Y=\left(-U(1)_1+3U(1)_5\right)/6$}}  
\\
\footnotesize
$X_{333}=p_1, X_{334}=p_2, X_{335}=p_{3}, X_{344}=1+p_1, X_{345}=p_4, X_{355}=p_5, X_{444}=p_6$,
$X_{445}=-1-3 p_3$, 
\\
\footnotesize
$X_{455}=p_7, X_{555}=0, X^\prime_{33}=p_{10}, X^\prime_{34}=p_{11}, X^\prime_{35}=p_{12}$, 
$X^\prime_{44}=p_{13},  X^\prime_{45}=p_{14}$, 
$X^\prime_{55}=p_8$, 
$X^{''}_{3}=p_{15}$,
\\
\footnotesize
$X^{''}_{4}=p_{16}$, $X^{''}_{5}=1+12 p_8$, $X^{'''}=p_9$, $Z_3=-8 p_1+12 p_3-6 p_5$,
\\
\footnotesize
$Z_4=-6 p_2-12 p_4-2 p_6-6 p_7$, 
$Z_5=0$, 
$Z^{'}=36+216 p_8-2 p_9$,
      \\ \hline
%32111
      \multicolumn{1}{|l|}{(xiii)$SO(6)\times SO(4)\times SO(2)^3\rightarrow G_{\rm SM}\times U(1)^4$}  
\\
\multicolumn{1}{|l|}{\underline{Hypercharge direction: $U(1)_Y=\left(-U(1)_1-3U(1)_3+3U(1)_4-3U(1)_5\right)/6$}}  
\\
See, Eq.~(\ref{eq:XZ32111}).
      \\ \hline
%431      
      \multicolumn{1}{|l|}{(xiv)-(a)$SO(8)\times SO(6)\times SO(2)\rightarrow SU(4)_C\times SU(3)\times U(1)^3\rightarrow G_{\rm SM}\times U(1)^4$}  
\\
\multicolumn{1}{|l|}{\underline{Hypercharge direction: $U(1)_Y=\left(U(1)_2-U(1)_3+3U(1)_4-U(1)_5\right)/6$}}  
\\
\footnotesize
$X_{333}=p_1, X_{334}=p_2, X_{335}=p_{3}, X_{344}=p_4, X_{345}=p_5, X_{355}=p_6, X_{444}=p_7$,
$X_{445}=p_8, X_{455}=p_9$, 
\\
\footnotesize
$X_{555}=p_{10}, X^\prime_{33}=p_1+p_2+2 p_{11}, X^\prime_{34}=p_1+p_4+p_{11}+2 p_{12}, 
X^\prime_{35}=p_{13}$, 
\\
\footnotesize
$X^\prime_{44}=14+8 p_1+4 p_2+4 p_3+7 p_4+15 p_7+4 p_8+12 p_{11}+4 p_{12}+8 p_{13}+16 p_{14}$,  
\\
\footnotesize
$X^\prime_{45}=5+2 p_1+p_2+p_3+p_4+3 p_5+4 p_7-3 p_8+12 p_9+3 p_{11}+p_{12}+2 p_{13}+4 p_{14}$, 
\\
\footnotesize
$X^\prime_{55}=2+2 p_1-2 p_2+7 p_3+p_4-3 p_5+9 p_6+p_7-p_8+3 p_9+4 p_{10}+p_{12}-p_{13}+p_{14}$, 
\\
\footnotesize
$X^{''}_{3}=p_{15}$, $X^{''}_{4}=p_{16}$, 
\\
\footnotesize
$X^{''}_{5}=22+28 p_1-34 p_2+101 p_3+14 p_4-48 p_5+96 p_6+8 p_7-7 p_8+24 p_9+16 p_{10}-6 p_{11}+14 p_{12}$
\\
\footnotesize
\hspace{25pt}$-14 p_{13}+8 p_{14}$, 
\\
\footnotesize
$X^{'''}=246+282 p_1-354 p_2+996 p_3+150 p_4-432 p_5+864 p_6+98 p_7-108 p_8+384 p_9+64 p_{10}-72 p_{11}$
\\
\footnotesize
\hspace{25pt}$+168 p_{12}-168 p_{13}+96 p_{14}+9 p_{15}-3 p_{16}$, 
\\
\footnotesize
$Z_3=-18 p_1+18 p_2-72 p_3-6 p_4+48 p_5-96 p_6$,
$Z_4=-2 p_7$, 
$Z_5=-6 p_8+24 p_9-32 p_{10}$, 
\\
\footnotesize
$Z^{'}=-54 p_1+54 p_2-216 p_3-18 p_4+144 p_5-288 p_6+2 p_7-24 p_8+96 p_9-128 p_{10}$,
      \\ \hline
%341      
      \multicolumn{1}{|l|}{(xiv)-(b)$SO(8)\times SO(6)\times SO(2)\rightarrow SU(4)_C\times SU(3)\times U(1)^3\rightarrow G_{\rm SM}\times U(1)^4$}  
\\
\multicolumn{1}{|l|}{\underline{Hypercharge direction: $U(1)_Y=\left(-2U(1)_1+3U(1)_2-U(1)_3-2U(1)_4+6U(1)_5\right)/12$}}  
\\
\footnotesize
$X_{333}=p_1, X_{334}=p_2, X_{335}=p_{3}, X_{344}=p_4, X_{345}=p_5, X_{355}=p_6, X_{444}=p_7$,
$X_{445}=2 p_8, X_{455}=p_9$, 
\\
\footnotesize
$X_{555}=2p_{10}, X^\prime_{33}=p_{11}, X^\prime_{34}=p_{12}, X^\prime_{35}=p_{13}$, 
$X^\prime_{44}=p_8+p_{10}+2 p_{14}$,  
\\
\footnotesize
$X^\prime_{45}=2 p_1+8 p_2+4 p_3+p_4+5 p_5+2 p_6+5 p_7+9 p_8+2 p_9+4 p_{11}+2 p_{12}+9 p_{13}+5 p_{14}+10 p_{15}$, 
\\
\footnotesize
$X^\prime_{55}=-30-32 p_1-72 p_3+24 p_4+72 p_5+2 p_6-8 p_7-32 p_8+2 p_9+13 p_{10}+16 p_{11}-16 p_{12}+8 p_{14}$, 
\\
\footnotesize
$X^{''}_{3}=21 p_1+12 p_2+17 p_3+21 p_4+12 p_5+21 p_6+5 p_7+20 p_8+10 p_{10}+4 p_{11}+19 p_{12}+4 p_{13}+20 p_{14}+25 p_{16}$,
\\
\footnotesize
$X^{''}_{4}=-42 p_1+18 p_2-59 p_3+21 p_4+27 p_5-17 p_6+p_7-36 p_8+4 p_9-p_{10}+16 p_{11}-33 p_{12}+11 p_{13}$
\\
\footnotesize
\hspace{25pt}$-7 p_{14}+15 p_{15}-25 p_{16}$, 
\\
\footnotesize
$X^{''}_{5}=-45-120 p_1+32 p_2-176 p_3+100 p_4+200 p_5+8 p_6-12 p_7-50 p_8+8 p_9+44 p_{10}+80 p_{11}$
\\
\footnotesize
\hspace{25pt}$-56 p_{12}+40 p_{13}+52 p_{14}+40 p_{15}$, 
\\
\footnotesize
$X^{'''}=90+98 p_1+204 p_2+84 p_3+114 p_4-102 p_5+24 p_6+98 p_7+33 p_8+24 p_9-34 p_{10}+90 p_{13}+90 p_{15}$, 
\\
\footnotesize
$Z_3=-14 p_1-6 p_3-12 p_4-6 p_6-2 p_7-3 p_8-3 p_{10}-6 p_{12}-6 p_{14}-6 p_{16}$,
\\
\footnotesize
$Z_4=12 p_1-12 p_2+18 p_3-6 p_4+6 p_5+6 p_6+18 p_8+6 p_{12}+6 p_{13}+6 p_{14}+6 p_{15}+6 p_{16}$, 
$Z_5=-4 p_{10}$, 
\\
\footnotesize
$Z^{'}=-4 p_1-24 p_2+24 p_3-36 p_4+12 p_5-4 p_7+30 p_8-4 p_{10}+12 p_{13}+12 p_{15}$,
      \\ \hline
%341.2      
      \multicolumn{1}{|l|}{(xiv)-(c)$SO(8)\times SO(6)\times SO(2)\rightarrow SU(4)_C\times SU(3)\times U(1)^3\rightarrow G_{\rm SM}\times U(1)^4$}  
\\
\multicolumn{1}{|l|}{\underline{Hypercharge direction: $U(1)_Y=\left(-2U(1)_1+3U(1)_2-3U(1)_3+6U(1)_5\right)/12$}}  
\\
\footnotesize
$X_{333}=p_1, X_{334}=p_2, X_{335}=p_{3}, X_{344}=p_4, X_{345}=p_5, X_{355}=p_6, X_{444}=p_7$,
$X_{445}=2 p_8, X_{455}=p_9$, 
\\
\footnotesize
$X_{555}=2p_{10}, X^\prime_{33}=p_{11}, X^\prime_{34}=p_{12}, X^\prime_{35}=p_{13}$, 
\\
\footnotesize
$X^\prime_{44}=8 p_1+14 p_3+2 p_4+16 p_6+19 p_8+5 p_{10}+16 p_{11}+2 p_{13}+20 p_{14}$,  
$X^\prime_{45}=p_{15}$, 
\\
\footnotesize
$X^\prime_{55}=-30+32 p_1+56 p_3+66 p_6+40 p_8+29 p_{10}+64 p_{11}+8 p_{13}+80 p_{14}$, 
\\
\footnotesize
$X^{''}_{3}=36 p_1+47 p_3+61 p_6+50 p_8+25 p_{10}+59 p_{11}+9 p_{13}+75 p_{14}$,
\\
\footnotesize
$X^{''}_{4}=21 p_2+17 p_5+21 p_9+4 p_{12}+4 p_{15}+25 p_{16}$, 
\\
\footnotesize
$X^{''}_{5}=-45+128 p_1+220 p_3+256 p_6+210 p_8+108 p_{10}+256 p_{11}+36 p_{13}+320 p_{14}$, 
\\
\footnotesize
$X^{'''}=90+8 p_1-96 p_3-66 p_6-45 p_8-34 p_{10}-90 p_{11}-90 p_{14}$, 
\\
\footnotesize
$Z_3=-8 p_1-6 p_6-3 p_8-3 p_{10}-6 p_{11}-6 p_{14}$,
$Z_4=-6 p_2-6 p_5-2 p_7-6 p_9-6 p_{16}$, 
\\
\footnotesize
$Z_5=-4 p_{10}$, 
$Z^{'}=-16 p_1-12 p_6-6 p_8-4 p_{10}-12 p_{11}-12 p_{14}$,
      \\ \hline     
%4211              
      \multicolumn{1}{|l|}{(xv)$SO(8)\times SO(4)\times SO(2)^2\rightarrow SU(4)\times SU(2)\times U(1)^3\rightarrow G_{\rm SM}\times U(1)^4$}  
\\
\multicolumn{1}{|l|}{\underline{Hypercharge direction: $U(1)_Y=\left(3U(1)_2+3U(1)_4-U(1)_5\right)/6$}}  
\\
\footnotesize
$X_{333}=p_1, X_{334}=p_{14}, X_{335}=p_{2}, X_{344}=p_3, X_{345}=p_4, X_{355}=p_5, X_{444}=p_6$,
$X_{445}=p_7, X_{455}=p_8$, 
\\
\footnotesize
$X_{555}=p_9,  X^\prime_{33}=6+p_1, X^\prime_{34}=p_{15}, X^\prime_{35}=p_{10}$, 
$X^\prime_{44}=2+p_3+4 p_{11}$,  
$X^\prime_{45}=2+p_4-2 p_7+8 p_8+p_{11}$, 
\\
\footnotesize
$X^\prime_{55}=p_{12}$, 
$X^{''}_{3}=p_{13}$,
$X^{''}_{4}=p_{16}$, 
$X^{''}_{5}=-p_2-8 p_5+4 p_7-16 p_8+2 p_{10}+8 p_{12}$, 
$X^{'''}=-18-2 p_1+3 p_{13}$, 
\\
\footnotesize
$Z_3=-2 p_1$,
$Z_4=-2 p_6$, 
$Z_5=-6 p_7+24 p_8-32 p_9$, 
$Z^{'}=-2 p_1$,
      \\ \hline      
%35           
       \multicolumn{1}{|l|}{(xvi)-(a)$SO(6)\times SO(10)\rightarrow SU(3)_C\times SU(5)\times U(1)^2\rightarrow G_{\rm SM}\times U(1)^4$}  
\\
\multicolumn{1}{|l|}{\underline{Hypercharge direction: $U(1)_Y=\left(-10U(1)_1+6U(1)_2+9U(1)_3-5U(1)_4-10U(1)_5\right)/60$}}  
\\
\footnotesize
$X_{333}=p_1, X_{334}=p_{2}, X_{335}=p_{3}, X_{344}=p_4, X_{345}=p_5, X_{355}=p_6, X_{444}=p_7$,
\\
\footnotesize
$X_{445}=1+p_1+p_2+p_3+p_4+p_7+2 p_8, X_{455}=2+2 p_1+2 p_2+2 p_3+2 p_4+4 p_8+5 p_9, X_{555}=p_{10}$,
\\
\footnotesize
$X^\prime_{33}=p_{11}, X^\prime_{34}=p_{12}, 
X^\prime_{35}=p_1+p_2+p_3+p_4+2 p_5+p_6+p_{11}+2 p_{12}+3 p_{13}$, 
\\
\footnotesize
$X^\prime_{44}=1+p_{11}+2 p_{14}$,  
$X^\prime_{45}=3+25 p_1+5 p_5+5 p_8+5 p_9-2 p_{11}+5 p_{12}+5 p_{13}+p_{14}$, 
\\
\footnotesize
$X^\prime_{55}=38+30 p_1+30 p_2+30 p_3+30 p_4+20 p_5+5 p_6+20 p_7+20 p_8+5 p_9+5 p_{10}+8 p_{11}+36 p_{14}+40 p_{15}$, 
\\
\footnotesize
$X^{''}_{3}=24+p_1+56 p_2+56 p_3+33 p_4-78 p_5-15 p_6+72 p_8+72 p_9-4 p_{11}+6 p_{12}+6 p_{13}-24 p_{14}-24 p_{15}$,
\\
\footnotesize
$X^{''}_{4}=4+6 p_1+3 p_2+4 p_3+2 p_4+5 p_7+2 p_8+5 p_9+6 p_{11}+4 p_{12}+2 p_{13}+6 p_{14}+8 p_{16}$, 
\\
\footnotesize
$X^{''}_{5}=269+459 p_1+137 p_2+136 p_3+193 p_4-60 p_5-30 p_6+95 p_7+328 p_8+250 p_9+25 p_{10}+14 p_{11}$
\\
\footnotesize
\hspace{25pt}$-34 p_{12}-32 p_{13}+114 p_{14}+120 p_{15}-8 p_{16}$, 
\\
\footnotesize
$X^{'''}=765-870 p_1+2280 p_2+2280 p_3-360 p_4-990 p_5-825 p_6+500 p_7+1710 p_8+1335 p_9+125 p_{10}$
\\
\footnotesize
\hspace{25pt}$-120 p_{11}-120 p_{12}-120 p_{13}-120 p_{14}-120 p_{15}$, 
\\
\footnotesize
$Z_3=-36-122 p_1-12 p_2-12 p_3-36 p_4+9 p_5-12 p_7-54 p_8-45 p_9-3 p_{10}+6 p_{11}-9 p_{12}-9 p_{13}$
\\
\footnotesize
\hspace{25pt}$-9 p_{14}-9 p_{15}$,
\\
\footnotesize
$Z_4=-51-96 p_1-48 p_2-39 p_3-42 p_4-15 p_5-23 p_7-57 p_8-60 p_9-5 p_{10}-3 p_{11}-9 p_{12}-12 p_{13}$
\\
\footnotesize
\hspace{25pt}$-30 p_{14}-30 p_{15}-3 p_{16}$, 
\\
\footnotesize
$Z_5=-174-351 p_1-99 p_2-108 p_3-120 p_4+15 p_5+15 p_6-69 p_7-222 p_8-150 p_9-18 p_{10}+3 p_{11}$
\\
\footnotesize
\hspace{25pt}$-6 p_{12}-3 p_{13}-75 p_{14}-75 p_{15}+3 p_{16}$, 
\\
\footnotesize
$Z^{'}=-585-405 p_1-555 p_2-555 p_3-270 p_4-135 p_5+75 p_6-280 p_7-585 p_8-375 p_9-70 p_{10}-90 p_{11}$
\\
\footnotesize
\hspace{25pt}$+60 p_{12}+60 p_{13}-390 p_{14}-390 p_{15}$,
      \\ \hline      
%35.2           
       \multicolumn{1}{|l|}{(xvi)-(b)$SO(6)\times SO(10)\rightarrow SU(3)_C\times SU(5)\times U(1)^2\rightarrow G_{\rm SM}\times U(1)^4$}  
\\
\multicolumn{1}{|l|}{\underline{Hypercharge direction: $U(1)_Y=\left(-10U(1)_1+6U(1)_2+9U(1)_3-5U(1)_4-10U(1)_5\right)/60$}}  
\\
\footnotesize
$X_{333}=p_1, X_{334}=p_{2}, X_{335}=p_{3}, X_{344}=p_4, X_{345}=p_5, X_{355}=p_6, X_{444}=p_7$,
$X_{445}=p_8,$ 
$X_{455}=4 p_7+5 p_9$, 
\\
\footnotesize
$X_{555}=p_{10}$,
$X^\prime_{33}=p_1+2 p_2+4 p_{11}, X^\prime_{34}=p_{12}, X^\prime_{35}=p_{13}$, 
$X^\prime_{44}=5+5 p_4+5 p_7+5 p_9+10 p_{14}$,  
$X^\prime_{45}=p_{15}$, 
\\
\footnotesize
$X^\prime_{55}=45+5 p_6+35 p_7+55 p_9+90 p_{14}$, 
$X^{''}_{3}=24+25 p_1-80 p_2+60 p_4+32 p_7+32 p_9-4 p_{12}+32 p_{14}$,
\\
\footnotesize
$X^{''}_{4}=-36-25 p_2-40 p_7-20 p_9+10 p_{12}-80 p_{14}$, 
$X^{''}_{5}=7 p_3+4 p_5+4 p_8+2 p_{13}+4 p_{15}+8 p_{16}$, 
\\
\footnotesize
$X^{'''}=-60+425 p_1-1350 p_2+900 p_4-820 p_7-420 p_9-300 p_{11}-60 p_{12}-120 p_{14}$, 
\\
\footnotesize
$Z_3=3-35 p_1+45 p_2-36 p_4-8 p_7-15 p_{11}+6 p_{12}+12 p_{14}$,
$Z_4=21+3 p_7+15 p_9+45 p_{14}$, 
\\
\footnotesize
$Z_5=-12 p_3-9 p_5-3 p_8-2 p_{10}+3 p_{13}-3 p_{16}$, 
\\
\footnotesize
$Z^{'}=-195-175 p_1+225 p_2-180 p_4-70 p_7-150 p_9-75 p_{11}+30 p_{12}-390 p_{14}$,
      \\ \hline  
%35.3           
       \multicolumn{1}{|l|}{(xvi)-(c)$SO(6)\times SO(10)\rightarrow SU(3)_C\times SU(5)\times U(1)^2\rightarrow G_{\rm SM}\times U(1)^4$}  
\\
\multicolumn{1}{|l|}{\underline{Hypercharge direction: $U(1)_Y=\left(-5U(1)_1+3U(1)_2+U(1)_3-5U(1)_4+15U(1)_5\right)/60$}}  
\\
\footnotesize
$X_{333}=p_1, X_{334}=p_{2}, X_{335}=p_{3}, X_{344}=p_4, X_{345}=p_5, X_{355}=p_6, X_{444}=p_7$,
$X_{445}=p_8, X_{455}=p_9$, 
\\
\footnotesize
$X_{555}=2 p_8+p_9+5 p_{10}$,
$X^\prime_{33}=p_{11}, X^\prime_{34}=p_{12}, 
X^\prime_{35}=p_1+p_2+p_5+p_6+p_{11}+2 p_{13}$, 
\\
\footnotesize
$X^\prime_{44}=3 p_4+p_7+3 p_8+4 p_{14}$,  
\\
\footnotesize
$X^\prime_{45}=5 p_1+8 p_4+5 p_5+5 p_6+6 p_7+3 p_8+5 p_{10}+5 p_{11}+5 p_{13}+4 p_{14}+10 p_{15}$, 
\\
\footnotesize
$X^\prime_{55}=5 p_1-5 p_2+25 p_3+16 p_4+15 p_6+12 p_7+6 p_8+15 p_{10}+5 p_{11}+8 p_{14}+20 p_{15}$, 
\\
\footnotesize
$X^{''}_{3}=5 p_1+6 p_3+7 p_4+5 p_6+4 p_{11}+2 p_{12}+4 p_{13}+8 p_{16}$,
\\
\footnotesize
$X^{''}_{4}=24+30 p_1-113 p_2+88 p_3+32 p_4-84 p_5+22 p_6+13 p_7-14 p_8+15 p_9+22 p_{10}-18 p_{11}-10 p_{12}$
\\
\footnotesize
\hspace{25pt}$-26 p_{13}+8 p_{14}-4 p_{15}-16 p_{16}$, 
\\
\footnotesize
$X^{''}_{5}=24+70 p_1-112 p_2-43 p_3+8 p_4-6 p_5-32 p_6+12 p_7-41 p_8+25 p_9-17 p_{10}-2 p_{11}+6 p_{13}-8 p_{14}$
\\
\footnotesize
\hspace{25pt}$+4 p_{15}-24 p_{16}$, 
\\
\footnotesize
$X^{'''}=-575 p_1+585 p_2-375 p_3-345 p_4+180 p_5-165 p_6+65 p_7-445 p_8+100 p_9-415 p_{10}-150 p_{11}-30 p_{12}$
\\
\footnotesize
\hspace{25pt}$-90 p_{13}-60 p_{14}-120 p_{15}$, 
\\
\footnotesize
$Z_3=-20 p_1+12 p_2-6 p_3-6 p_4+9 p_5-3 p_6+2 p_7+6 p_{11}-3 p_{12}+3 p_{13}+6 p_{15}-3 p_{16}$,
\\
\footnotesize
$Z_4=-9+33 p_2-33 p_3+54 p_5+3 p_6+3 p_7+12 p_8-9 p_9+3 p_{10}+18 p_{11}+21 p_{13}+3 p_{14}+24 p_{15}+6 p_{16}$, 
\\
\footnotesize
$Z_5=-9-15 p_1+42 p_2-12 p_3+15 p_4+36 p_5+12 p_6+9 p_7+8 p_8-8 p_9+2 p_{10}+12 p_{11}+9 p_{13}+12 p_{14}$
\\
\footnotesize
\hspace{25pt}$+21 p_{15}+9 p_{16}$, 
\\
\footnotesize
$Z^{'}=175 p_1-105 p_2-75 p_3-45 p_4+45 p_5-30 p_6-40 p_7+20 p_8-5 p_9+5 p_{10}+15 p_{12}+45 p_{13}-45 p_{14}-15 p_{15}$,
      \\ \hline 
%3221           
       \multicolumn{1}{|l|}{(xvii)$SO(6)\times SO(4)\times SO(4)\times SO(2)\rightarrow SU(3)_C\times SU(2)_L\times SU(2)\times U(1)^4\rightarrow G_{\rm SM}\times U(1)^4$}  
\\
\multicolumn{1}{|l|}{\underline{Hypercharge direction: $U(1)_Y=\left(-U(1)_1+3U(1)_3+3U(1)_4\right)/6$}}  
\\
\footnotesize
$X_{333}=p_1, X_{334}=p_{2}, X_{335}=p_{3}, X_{344}=p_4, X_{345}=p_{11}, X_{355}=p_1+2 p_2+p_4, X_{444}=-8 p_1-12 p_2-6 p_4$,
\\
\footnotesize
$X_{445}=-3+p_2$, 
\\
\footnotesize
$X_{455}=p_{12}, X_{555}=p_5$,
$X^\prime_{33}=p_6, X^\prime_{34}=p_7, X^\prime_{35}=p_{13}$, 
$X^\prime_{44}=p_8$,  
$X^\prime_{45}=p_{14}$, 
$X^\prime_{55}=p_{15}$, 
$X^{''}_{3}=p_9$,
\\
\footnotesize
$X^{''}_{4}=3+16 p_6+16 p_7+4 p_8-2 p_9$, 
$X^{''}_{5}= p_{16}$, 
$X^{'''}=p_{10}$, 
\\
\footnotesize
$Z_3=-8 p_1-12 p_2-6 p_4$,
$Z_4=16 p_1+24 p_2+12 p_4$, 
$Z_5=-6 p_3-2 p_5$, 
$Z^{'}=36+96 p_6+96 p_7+24 p_8-2 p_{10}$,
      \\ \hline 
%3311           
              \multicolumn{1}{|l|}{(xviii)$SO(6)\times SO(6)\times SO(2)^2\rightarrow SU(3)_C\times SU(3)\times U(1)^4\rightarrow G_{\rm SM}\times U(1)^4$}  
\\
\multicolumn{1}{|l|}{\underline{Hypercharge direction: $U(1)_Y=\left(-U(1)_1+3U(1)_2-U(1)_3+U(1)_4+3U(1)_5\right)/6$}}  
\\
\footnotesize
$X_{333}=p_1, X_{334}=p_{2}, X_{335}=p_{3}, X_{344}=p_4, X_{345}=p_5, X_{355}=p_6, X_{444}=p_7$,
$X_{445}=p_8, X_{455}=p_9$, 
\\
\footnotesize
$X_{555}=p_{10}$,
$X^\prime_{33}=p_{11}, X^\prime_{34}=p_{12}, X^\prime_{35}=p_{13}$, 
$X^\prime_{44}=6+p_1+3 p_2+3 p_4+p_7-p_{11}-2 p_{12}$,  
\\
\footnotesize
$X^\prime_{45}=6+6 p_3-9 p_5-4 p_6+3 p_8+2 p_9+2 p_{13}$, 
$X^\prime_{55}=p_{14}$, 
$X^{''}_{3}=p_{15}$,
$X^{''}_{4}=p_{16}$, 
\\
\footnotesize
$X^{''}_{5}= 60-2 p_1+15 p_2+27 p_3+18 p_4-36 p_5-18 p_6+p_7+9 p_8+6 p_9-18 p_{12}+6 p_{13}+2 p_{14}+2 p_{15}-p_{16}$, 
\\
\footnotesize
$X^{'''}=-18-2 p_1-6 p_2-6 p_4-2 p_7+3 p_{15}+3 p_{16}$, 
$Z_3=-6 p_1+6 p_2-6 p_4$,
\\
\footnotesize
$Z_4=4 p_1-12 p_2-2 p_7$, 
$Z_5=-2 p_{10}$, 
$Z^{'}=-2 p_1-6 p_2-6 p_4-2 p_7$,
      \\ \hline                  
  \caption{The list of 24 variables $\{X_{ABC}, X^\prime_{AB}, X^{''}_A, X^{'''}, Z_A, Z^\prime\}$ with integers $p_m$, leading to three-generation models on simply-connected CY threefolds. We now choose a specific hypercharge direction for simplicity.}
  \label{tab:hflux}
\end{longtable}

\end{document}